\documentclass{JINST}
\usepackage{amsmath,amssymb,mathrsfs,graphicx}

\title{Study of a high spatial resolution $^{10}B$-based thermal neutron
detector for application in neutron reflectometry: the Multi-Blade prototype}

\author{F. Piscitelli$^{a,b}$\thanks{Corresponding
author.},  J.C. Buffet$^a$, J.F. Clergeau$^a$, S. Cuccaro$^a$, B. Gu\'erard$^a$, A. Khaplanov$^{a,c}$, Q. La Manna$^a$, J.M. Rigal$^a$ and P. Van Esch$^a$\\
\llap{$^a$} Institut Laue-Langevin (ILL),\\ 6, Jules Horowitz, 38042
Grenoble, France. \\
\llap{$^b$} Department of Physics, University of Perugia,\\
Piazza Universit\`a 1, 06123 Perugia, Italy.\\
\llap{$^c$} European Spallation Source, \\
P.O. Box 176, SE-22100 Lund, Sweden.\\
E-mail: \email{piscitelli@ill.fr}}

\abstract{Although for large area detectors it is crucial to find an
alternative to detect thermal neutrons because of the $^3He$
shortage, this is not the case for small area detectors. Neutron
scattering science is still growing its instruments' power and the
neutron flux a detector must tolerate is increasing. For small area
detectors the main effort is to expand the detectors' performances.
\\ At Institut Laue-Langevin (ILL) we developed the Multi-Blade
detector which wants to increase the spatial resolution of
$^3He$-based detectors for high flux applications. We developed a
high spatial resolution prototype suitable for neutron reflectometry
instruments. It exploits solid $^{10}B$-films employed in a
proportional gas chamber. Two prototypes have been constructed at
ILL and the results obtained on our monochromatic test beam line are
presented here.}


\begin{document}
\section{Introduction}
Although the $^3He$ shortage pushes the research of alternatives for
large area detectors, this is not the case for small area detectors,
where the main effort is to expand the detectors' performances.
\\ At Institut Laue-Langevin (ILL) in Grenoble (France) we developed the Multi-Blade detector
which wants to push the limit of spatial resolution of $^3He$-based
detectors for high flux applications. The Multi-Blade was conceived
to be a detector suitable for neutron reflectometry instruments
which require a high spatial resolution.
\\ Moreover, neutron scattering science is still growing its
instruments' power and the neutron flux a detector must tolerate is
increasing. The peak brightness at the new European Spallation
Source (ESS) in Lund (Sweden) will be higher than that of any of the
short pulse sources, and will be more than one order of magnitude
higher than that of the world's leading continuous source. The
time-integrated brightness at ESS will also be one to two orders of
magnitude larger than is the case at today's leading pulsed sources
\cite{esstdr}, \cite{gebauer1}. The Multi-Blade can also address the
counting rate capability of $^3He$ detectors looking forward higher
fluxes.
\\ A neutron detector for a reflectometer is in general compact in
size (about $400\times250\,mm^2$ \cite{figaro}) and it is
characterized by a non-uniform spatial resolution. In order to
achieve the needed angular resolution a high spatial resolution is
only needed in one direction and it is of the order of $1\,mm$
\cite{cubittD17}. Such a resolution is really needed only for
off-specular studies. To be more precise a Position Sensitive
Detector (PSD) is necessary when not only specular reflection occurs
but one wants to quantify more sample features, e.g. off-specular
reflection arising from the presence of in-plane structures.
\\ For a large number of applications only the specular reflection is
needed and the other detector coordinate is generally integrated
over. The $1\,mm$ spatial resolution is still feasible with $^3He$
detector but in many areas of soft and hard matter research science,
the amount of material to investigate is rather limited. Partly
because the fabrication of larger samples is too expensive or not
feasible, yet, partly because the interesting features depend on the
size. The development of a neutron reflectometer optimized for small
samples is under study \cite{rainbow1}. There is a great deal of
interest in expanding the technique of neutron reflectometry beyond
static structural measurements of layered structures to kinetic
studies \cite{cubitt2}. The time resolution for kinetic studies is
limited by the available neutron flux.
\\ For a reflectometer the actual neutron flux reaching the sample is a small fraction of
the incoming neutron beam. In \cite{cubitt2} and \cite{rainbow2} a
new instrument layout is presented to open the possibility of
sub-second kinetic studies, however this requires very high spatial
resolution detectors. The wanted resolution, well beyond the
reasonable limit of the $^3He$ technology, is $\Delta x=0.2\,mm$;
required in one dimension only (the other dimension can be summed)
\cite{cubitt2}.
\\ CCD cameras instead of gaseous detectors con be used to improve the spatial resolution but
Time-of-Flight measurements are not possible.
\\ Moreover, $^3He$-based detectors are also limited in counting rate capability by the
space charge effect.
\\ The Multi-Blade detector represents a promising
alternative to $^{3}He$-based detectors, to accomplish the high
spatial resolution and the high count rate capability needed in the
new applications of neutron reflectometry. This alternative exploits
solid $^{10}B$-films employed in a proportional gas chamber. The
challenge with this technique is to attain a suitable detection
efficiency which is about $63\%$ for the Figaro \cite{figaro}
detector at $2.5$\AA. A suitable detection efficiency can be
achieved by operating the $^{10}B$ conversion layer at grazing angle
relative to the incoming neutron direction. The Multi-Blade design
is based on this operational principle and it is conceived to be
modular in order to be adaptable to different applications. Two
prototypes have been developed at ILL and the results obtained on
our monochromatic test beam line are presented here. A significant
concern in a modular design is the uniformity of detector response.
Several effects might contribute to degrade the uniformity and they
have to be taken into account in the detector concept: overlap
between different substrates, coating uniformity, substrate
flatness, etc.

\section{The Multi-Blade concept}
\begin{figure}[!ht]
\centering
\includegraphics[width=10cm,angle=0,keepaspectratio]{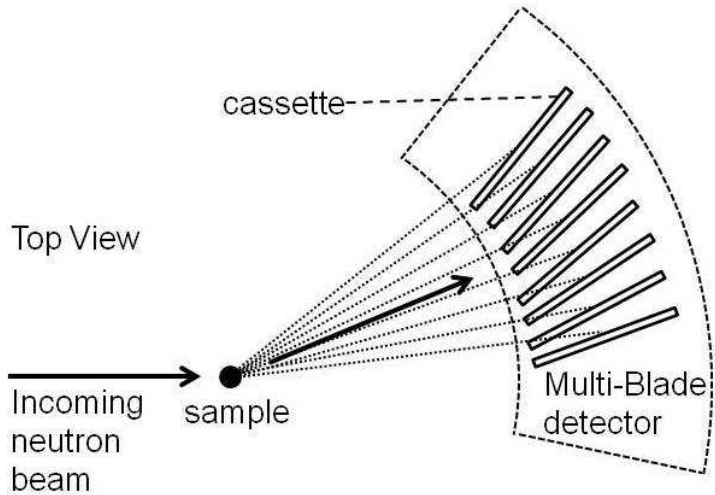}
\caption{\footnotesize The Multi-Blade detector sketch (top
view).}\label{schemMBlight6}
\end{figure}

\begin{figure}[!ht]
\centering
\includegraphics[width=4.8cm,angle=0,keepaspectratio]{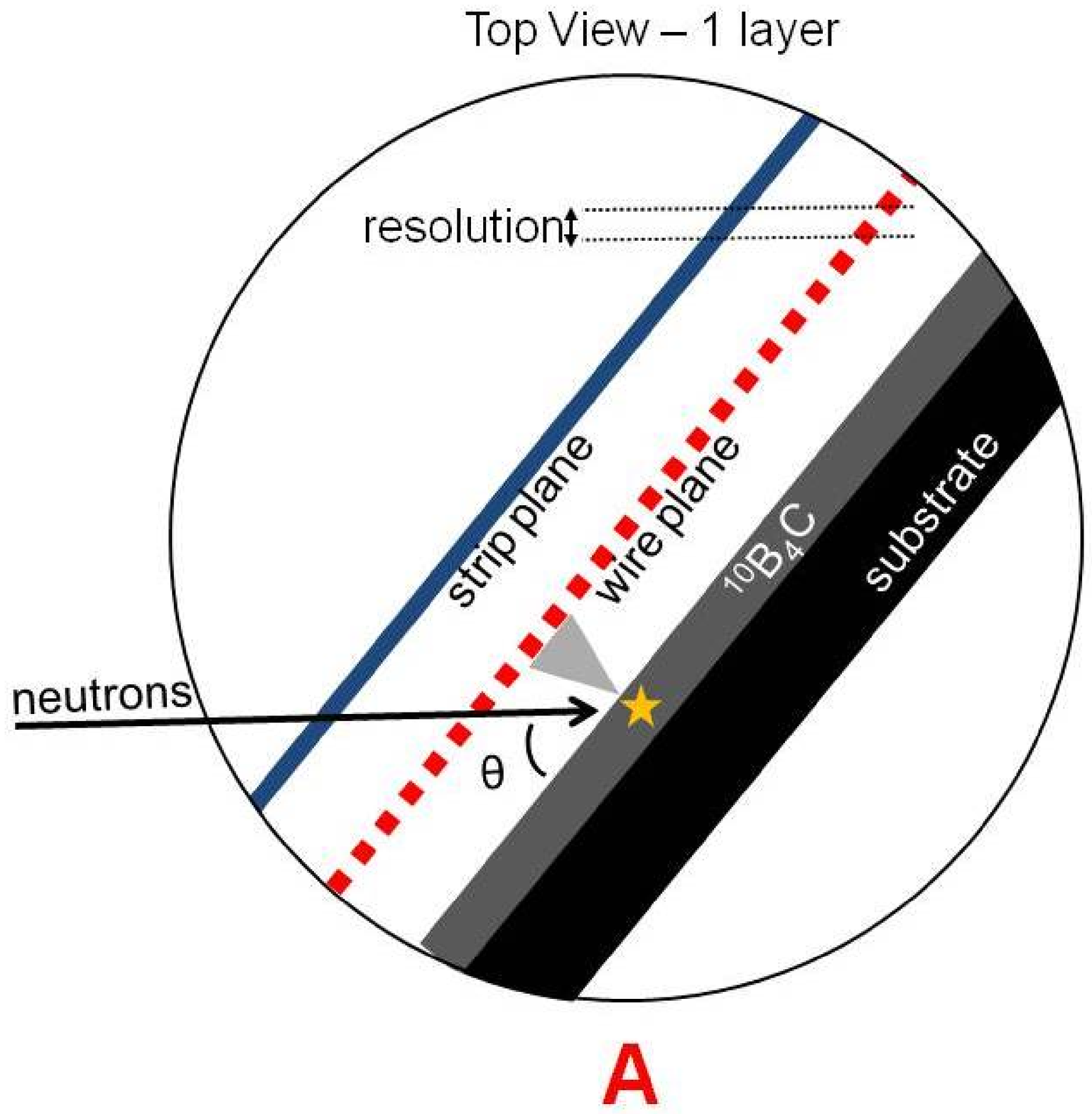}
\includegraphics[width=4.8cm,angle=0,keepaspectratio]{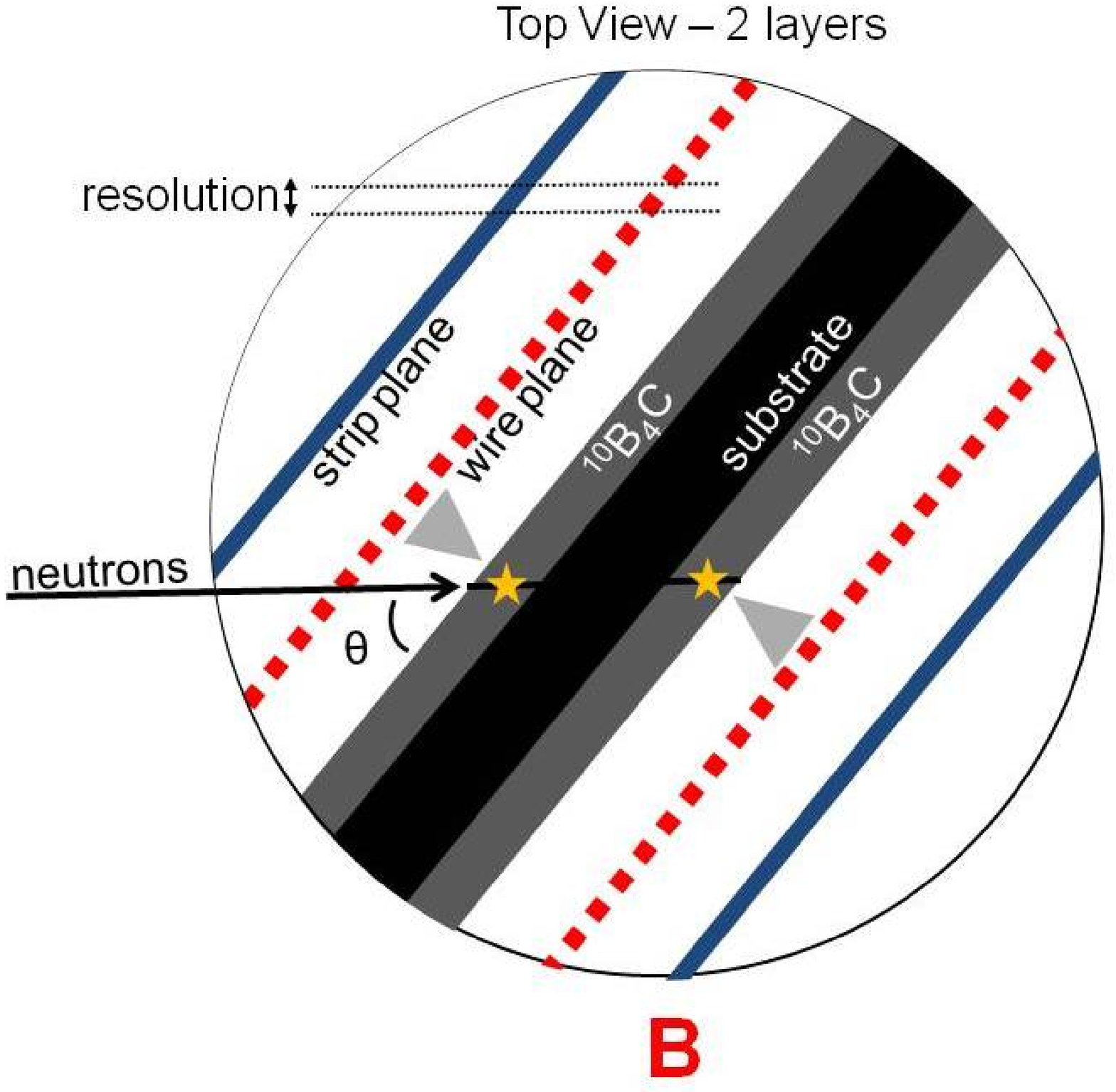}
\includegraphics[width=5.1cm,angle=0,keepaspectratio]{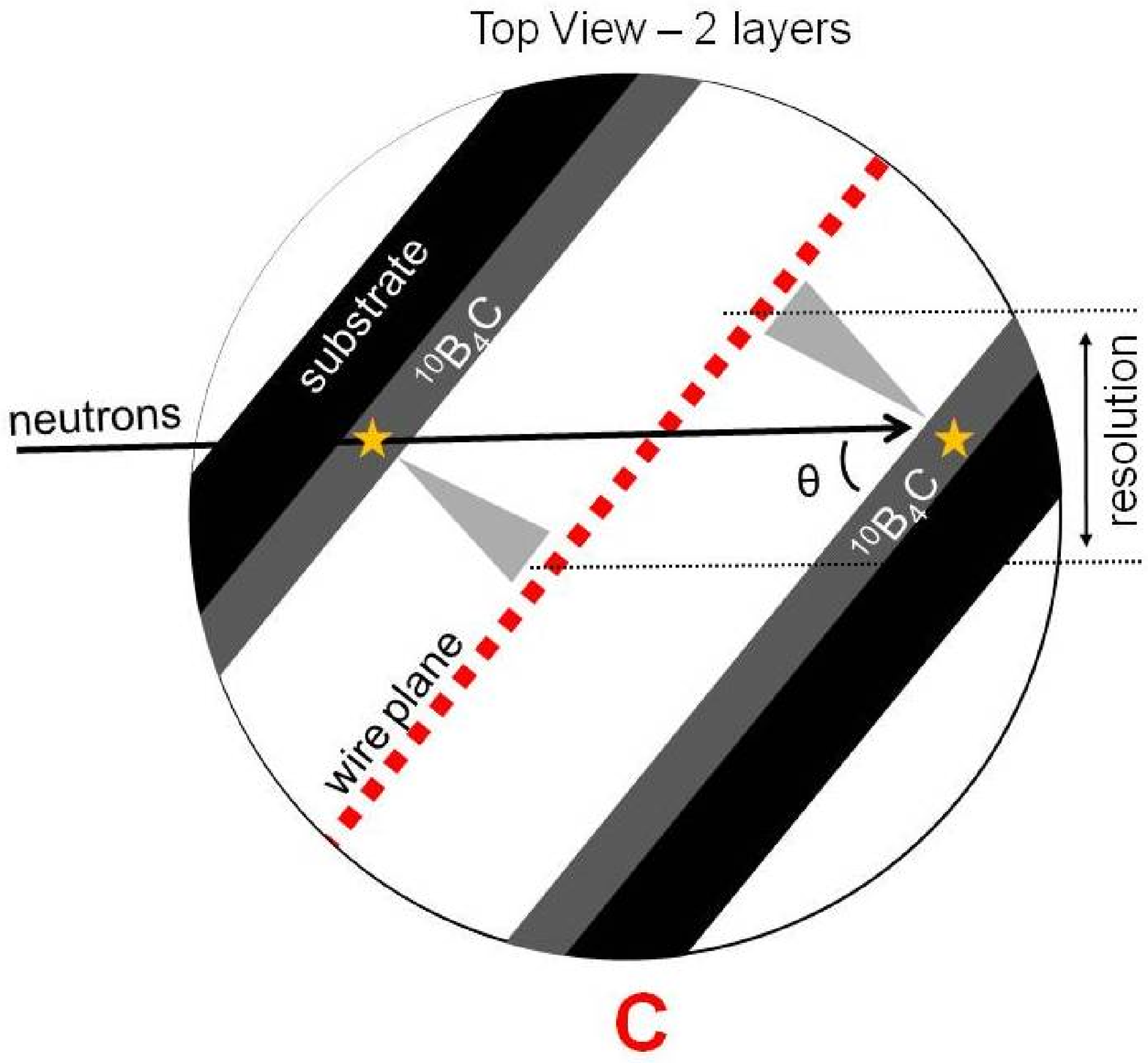}
\caption{\footnotesize Cross-section of one cassette. Three options
are shown: A, a single converter layer; B and C, with two
converters.}\label{figabc09}
\end{figure}

\begin{figure}[!ht]
\centering
\includegraphics[width=7.5cm,angle=0,keepaspectratio]{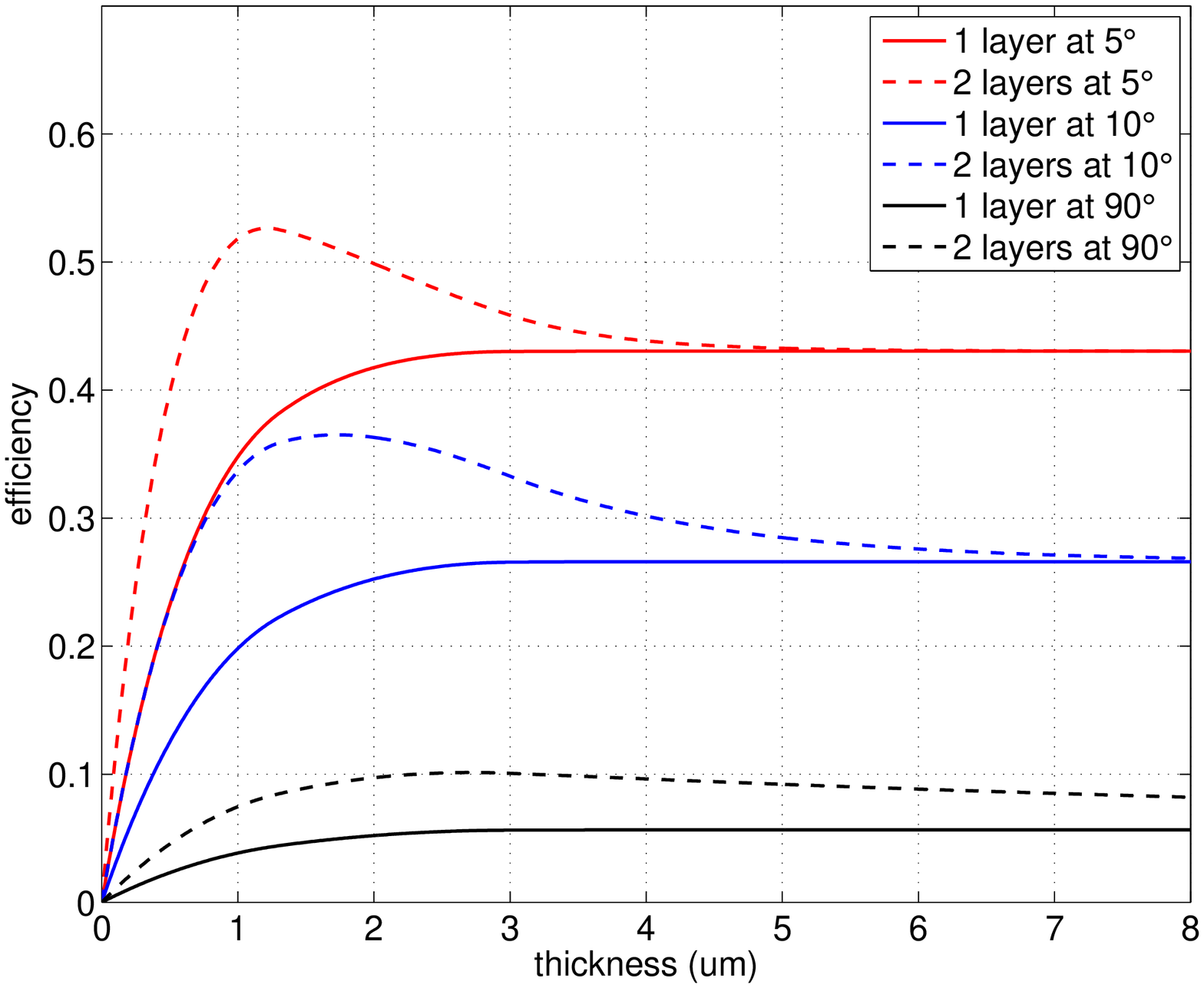}
\includegraphics[width=7.5cm,angle=0,keepaspectratio]{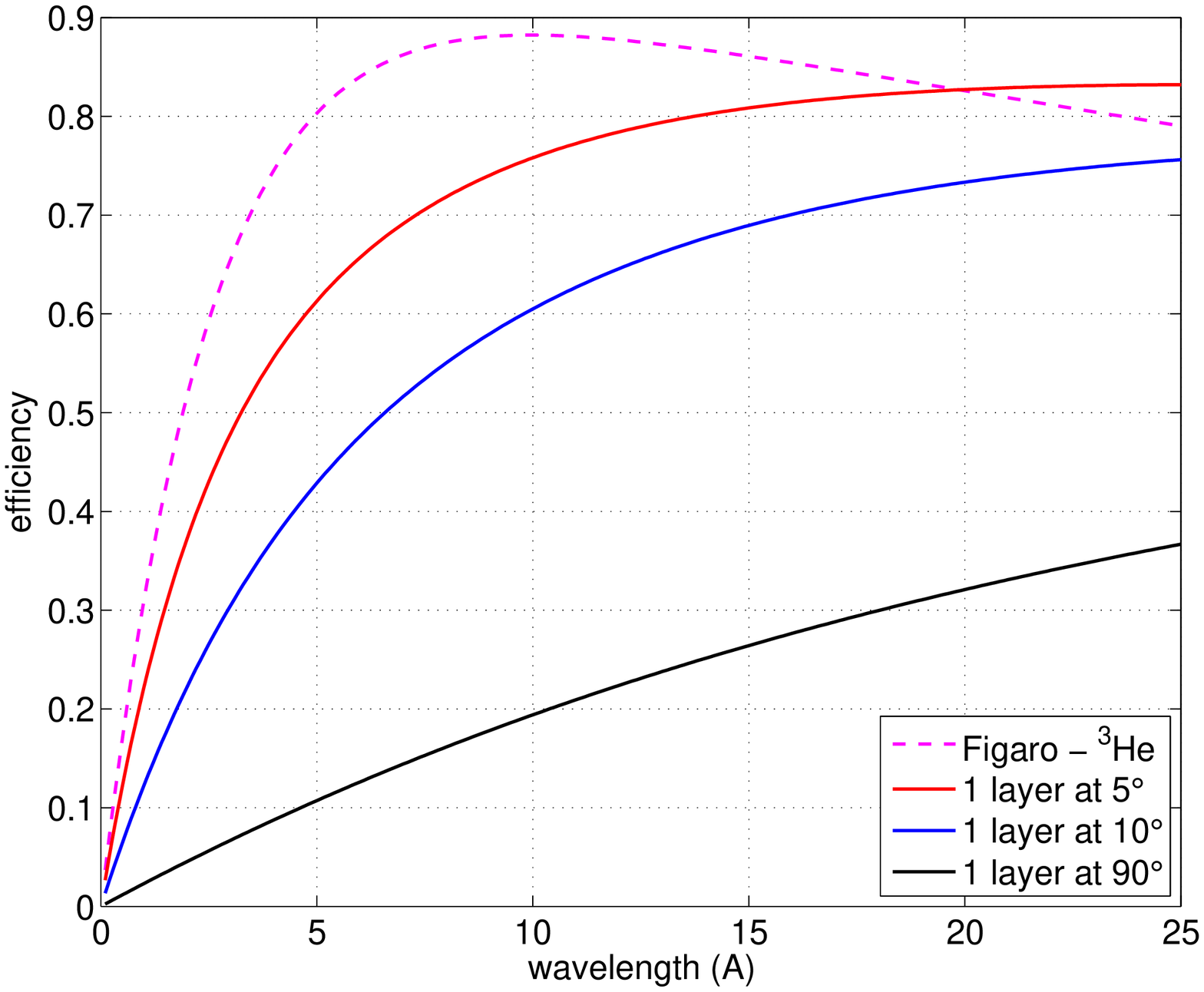}
\caption{\footnotesize $^{10}B_4C$ layers ($\rho=2.24\,g/cm^3$)
detection efficiency at $2.5$\AA \, as a function of the layer
thickness for the options A and B for three inclinations (left),
efficiency as a function of the neutron wavelength for three
inclinations of a single $3\,mu m$ layer (right). An energy
threshold of $100\,KeV$ is applied. The efficiency of the Figaro's
detector \cite{figaro} is also shown ($6.9\,mm$ tubes filled with
$8\,bars$ of $^3He$).}\label{figgig345htyigv}
\end{figure}

The Multi-Blade concept was already introduced at ILL in 2005
\cite{buff1} and a first prototype was realized in 2012
\cite{buff3}. Its design is conceived to be modular in order to be
versatile to be applied in many applications on several instruments.
The Multi-Blade exploits solid $^{10}B$-films employed as a neutron
converter in a proportional gas chamber as in \cite{jonisorma}. The
challenge with this technique is to attain a suitable detection
efficiency. This latter can be achieved by operating the $^{10}B$
conversion layer at grazing angle relative to the incoming neutrons
direction. Moreover the inclined geometry leads to a gain in spatial
resolution and as well in counting rate capability compared to
$^{3}He$ detectors.
\\ Figure \ref{schemMBlight6} shows the Multi-Blade detector
schematic, it is made up of several identical units called
\emph{cassettes}. Each \emph{cassette} acts as an independent Multi
Wire Proportional Chamber (MWPC) which holds both the neutron
converter and the read-out system. The fully assembled detector is
composed of several cassettes inclined toward the sample position.
The angle subtended by each cassette looking at the sample position
is kept constant in order to maintain the spatial resolution and the
efficiency as uniform as possible.
\\ The cassettes must be arranged taking into account an overlap
between them in order to avoid dead space over the whole detector
surface. If the instrument geometry changes the cassette arrangement
in the detector should also change.
\\ Each cassette should contain one or more neutron converters, e.g.
$^{10}B_4C$ layers, and the read-out system that has to assure the
two-dimensional identification of the neutron event. Figure
\ref{figabc09} shows the cross-section of the cassette concept for
three different configurations.
\\ In the A and B solutions, in each cassette a single converter layer
is facing each read-out system. The read-out is a wire plane and a
strip plane placed orthogonally. The space between the strips and
the converter is filled with stopping gas at atmospheric pressure to
ensure the gas multiplication. The converter layer is polarized as
well and acts as a cathode together with the strip plane; the wire
plane, on the other hand, acts as an anode plane.
\\ In the C configuration a single wire plane performs the
two-dimensional read-out through charge division on resistive wires.
A single read-out system is facing two converter layers. The space
between the two converters is filled with stopping gas. The two
converter layers act as cathodes.
\\ Referring to configurations A and B, the identification of the
position of a neutron event is the coincidence of wire and strip
hits. The spatial resolution given by the strips does not depend on
the inclination of the cassette. The spatial resolution given by the
wire plane improves as the angle with the incoming neutron direction
decreases. Note that if the detector is able to distinguish two
events hitting on two consecutive wires, the spatial resolution is
given by the wire pitch. Thanks to the inclination of the wire plane
with respect to the incoming beam, this pitch is diminished by a
factor $\sin(\theta)$ with $\theta$ the layer inclination. Hence,
the spatial resolution is improved by the same factor. E.g. at
$\theta=5^{\circ}$, the resolution is improved by a factor about
$10$ ($1/\sin(\theta=5^{\circ})\sim 10$).
\\ Moreover, the actual neutron flux over the detector would be
divided by the same factor increasing its counting rate capability;
the same flux is shared by several wires.
\\ While the spatial resolution is improved by inclination in both options A and B in Figure
\ref{figabc09}, all the advantage of working at grazing angle is
lost in the C configuration. In \cite{kleinjalousie} can be found
the actual implementation of such a detector. In options A and B the
charge generated by neutron capture fragments in the gas gives a
signal on the facing wires and strips. In the solution C, for a
given incoming neutron direction there will be two regions on the
converters where neutrons are converted. The smaller the angle at
which we operate the detector, the larger is the distance between
those two regions. The uncertainty on the conversion point is then
given by this distance which is much bigger than the wire pitch. On
the other hand, option C has half the number of read-out channels as
compared to A and B. However, for us, high spatial resolution is
crucial.
\\ We decided to concentrate on the implementation of the option A
and B.
\\ In \cite{buff3} has been derived how the solid converter layer
efficiency increases as a function of its inclination. We decided to
operate the Multi-Blade at either $\theta=10^{\circ}$ or
$\theta=5^{\circ}$ in order to have significant detection efficiency
and keeping the mechanics simple.
\\ Figure \ref{figgig345htyigv} shows the detection efficiency for $^{10}B_4C$
layers ($\rho=2.24\,g/cm^3$) calculated according to \cite{gregor}
and \cite{fratheo}. An energy threshold of $100\,KeV$ is applied. We
considered two possible configurations: options A and B in Figure
\ref{figabc09}, with one converter or two. On the left we show the
neutron detection efficiency, at $2.5$\AA, as a function of the
converter layer thickness for the solutions A and B. While the
efficiency shows a maximum for the two layer option, it is saturated
over $3\,\mu m$ for the single layer. The single converter option
can attain a maximum efficiency of $28\%$ at $10^{\circ}$ and $44\%$
at $5^{\circ}$ ($2.5$\AA) to be compared with the double-layer
efficiency of $37\%$ and $54\%$ respectively. The addition of the
second layer, at $\theta=5^{\circ}$ leads to an increase of the
efficiency of about $10\%$ with respect to the solution A. The
advantage of having only one converter is that the coating can be of
any thickness above $3\,\mu m$ and the efficiency is not affected,
while for the two layer option its thickness should be well
calibrated. Moreover, in the two layer configuration the substrate
choice is also crucial because it should be kept as thin as possible
to avoid neutron scattering and this leads to possible mechanical
issues. On the solution A, the substrate choice can be more flexible
because it has not to be crossed by neutrons.
\\ On the right in Figure \ref{figgig345htyigv} we show the efficiency as a
function of the neutron wavelength for the the single layer of
thickness $3\,\mu m$ (configuration A). The Figaro's detector
efficiency \cite{figaro} efficiency is also plotted, it is a
$^3He$-based detector made up of $6.9\,mm$ tubes filled at
$8\,bars$. In the plots shown the detector gas vessel Aluminium
window is also taken into account as a neutron loss. For the
Figaro's detector we used a $5\,mm$ thick window, and, since the
Multi-Blade detector will be operated at atmospheric pressure, we
used a $2\,mm$ window. $^3He$-based detectors' efficiency can be
increased by increasing the $^3He$ pressure in the vessel; on the
other hand, for a solid converter based detector the gas acts only
as a stopping means, hence its pressure can be kept at atmospheric
values. Consequently the gas vessel construction has less
constraints.
\\ In each of the solutions proposed for the cassette concept, see Figure \ref{figabc09},
the read-out system has to be crossed by neutrons before reaching
the converter. The mechanical challenge in the read-out system
construction is to minimize the amount of material on the neutron
path to avoid scattering that can cause misaddressed events in the
detector.
\\ The cassettes must overlap to avoid dead spaces and the event loss, due to the zone
where we switch the cassette, should be minimized. At the cassette
edge electric field distortions and structure holding materials can
cause a loss in the efficiency and consequently deteriorate the
detector uniformity. In the prototype realization all these problems
have been taken into account, two prototypes have been built in
order to study the possible issues.

\section{Multi-Blade version V1}\label{sectv1mbg}
\subsection{Mechanical study}
\begin{figure}[!ht]
\centering
\includegraphics[width=14cm,angle=0,keepaspectratio]{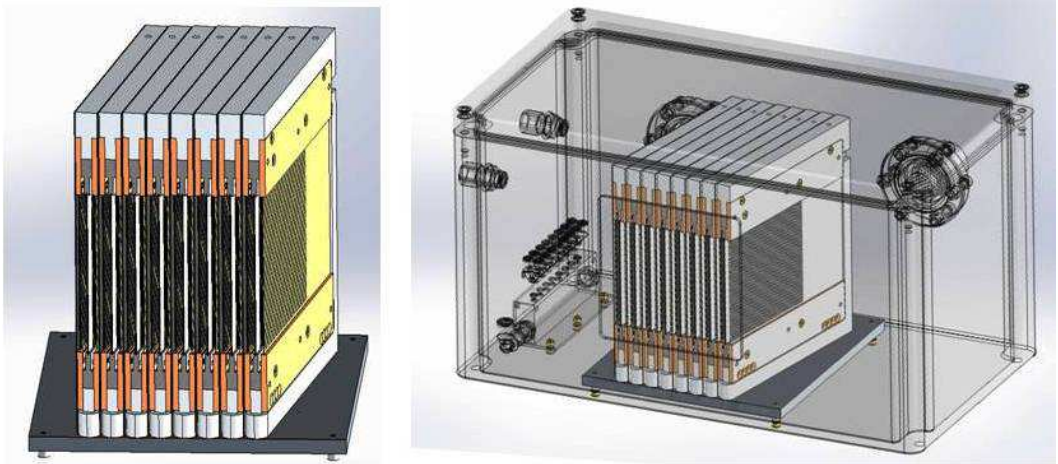}
\caption{\footnotesize A 8-cassettes Multi-Blade in its gas vessel.}
\label{dgfag5465yb5hv}
\end{figure}

The prototype was conceived to clarify the advantages and
disadvantages of the options A and B shown in Figure \ref{figabc09}.
\\ The detector works as a standard MWPC operated at atmospheric pressure
stopping gas such as $Ar/CO_2$ $(90/10)$.
\\ Since we want to avoid neutrons to be scattered before reaching the
converter layer, we need to minimize the amount of matter that has
to be crossed by neutrons: the read-out system and the cassette
window.
\begin{figure}[!ht]
\centering
\includegraphics[width=14cm,angle=0,keepaspectratio]{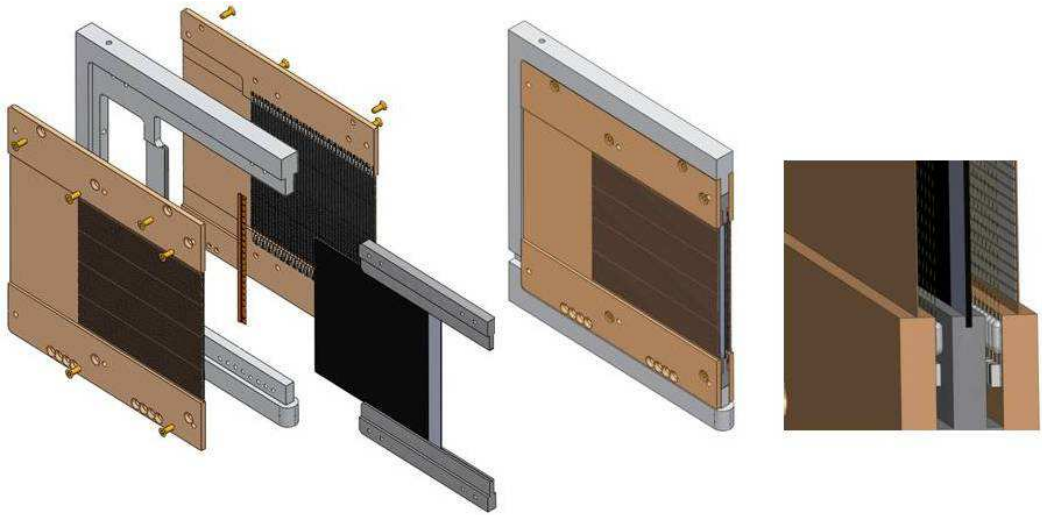}
\caption{\footnotesize Exploded and assembled view of a cassette
(left and center). Detail of a cassette (right): the two polyimide
PCBs surrounding the coated blade.} \label{viewfogi9fo}
\end{figure}
\\ Figure \ref{viewfogi9fo} shows a cassette drawing. An Aluminium
substrate of thickness $0.5\,mm$ is coated on both sides by a
$^{10}B_4C$-layer, i.e. a blade. One layer will work as a
back-scattering layer and the second as a transmission layer
\cite{fratheo}. The converter is surrounded symmetrically by two
polyimide PCBs. Each of them holds a cathode strip plane and a anode
wire plane. The converter layer substrate is grounded and it acts as
a cathode plane. Therefore a half cassette is a complete MWPC
containing one neutron converter layer and a two-dimensional
read-out system.
\\ The entire structure is supported by an Aluminium U-shaped holder. The latter shape was
conceived to remove any material that can scatter the incoming
neutrons. Moreover, each holder presents two gas inlets in order to
supply the stopping gas directly inside the gap between the
converter and the PCBs. The exhausted gas will flow out from the
frontal opening of the cassette inside the gas vessel.
\\ Figure \ref{viewfogi9fo} also shows the detail of an assembled
cassette. The polyimide PCBs have to be crossed before neutrons can
be converted, hence, in order to reduce the amount of material that
can induce neutron scattering, and thus misaddressed detected
events, those PCBs are as thin as possible according to the
mechanical constraints in their inner active region. The strips are
deposited on the polyimide and the anode wires are stretched
orthogonally over the strip plane. The copper strips are $0.8\, mm$
wide and spaced by $0.2\, mm$; tungsten wires are $15\, \mu m$ thick
and they are spaced by  $2.5\, mm$. The final electric signal is
obtained by gas amplification on the anode wires placed in the gas
volume. In order to decrease the number of read-out channels, anode
wires and cathode strips are grouped by resistive chain for charge
division read-out. Each full \emph{cassette} has then 4 anode
outputs and 4 cathodes outputs making 4 charge division read-out
chains. The resistors are placed on the PCBs surface. This readout
technique is cost-effective, but it is not the most performing
concerning the count rate. If we want to take advantage of the high
count rate capability a better readout technique must be used.
\\ The polyimide PCBs are $60\, \mu m$ thick in the inner region:
$25\, \mu m$ is the polyimide thickness and $35\, \mu m$ is the
copper strips thickness.
\\ The sensitive area of each cassette is $10\times 9\,cm^2$ but, since it will be oriented at $10\,^{\circ}$ with
respect to the incoming neutron direction, the actual sensitive area
offered to the sample is given by $(10\,cm \cdot
\sin(10\,^{\circ}))\times 9\,cm = 1.7\times 9\, cm^2$. As a result,
the actual wire pitch, at $10\,^{\circ}$, is improved down to
$0.43\,mm$.
\\ The detector will be installed to have the
better resolution in the direction of the reflectometry instrument
collimation slits; i.e. the cassettes, which can be mounted either
horizontally or vertically, will be oriented with the wires parallel
to the instrument slits.
\\ Figure \ref{dgfag5465yb5hv} shows a drawing of 8 cassettes stacked one after
the other and placed in the gas vessel.
\\ As already mentioned, the main issue to be addressed in the final
detector is the uniformity, as soon as it is made of several units,
their arrangement is crucial to get a uniform response in
efficiency. A misalignment in one of the modules can give rise to a
drop in the efficiency or dead zones.
\\ The cassettes have to be arranged in order to overlap to avoid
dead zones. For this reason this detector is suitable for fixed
geometry reflectometry instruments, where the distance between
sample and detector is kept constant and the arrangement does not
change.
\\ The final prototype will be mounted in a gas vessel together
with the gas distribution unit which splits the inlet in the several
cassettes, and the electronic connections. This is illustrated in
Figure \ref{dgfag5465yb5hv}.
\\ Since the gas is flushed cost effective materials can be used
because their outgassing is not an issue.

\subsection{Mechanics}
\begin{figure}[!ht]
\centering
\includegraphics[width=10cm,angle=0,keepaspectratio]{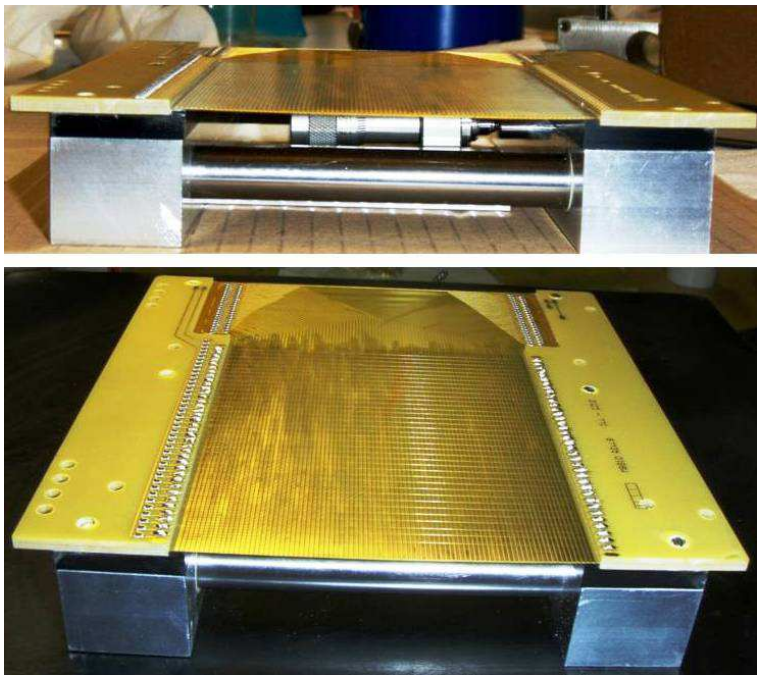}
\caption{\footnotesize A polyimide PCBs where anode wires are
mounted orthogonal to the cathode strips.} \label{figMBv1prima1}
\end{figure}

\begin{figure}[!ht]
\centering
\includegraphics[width=12cm,angle=0,keepaspectratio]{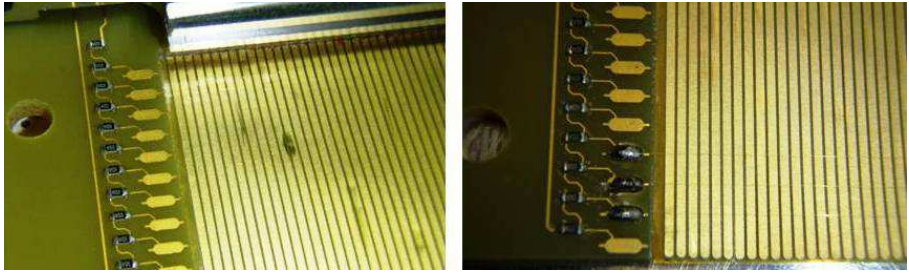}
\caption{\footnotesize Detail of a polyimide PCBs: resistors for
charge division link the wire pads where anodes are soldered.}
\label{figMBv1prima2}
\end{figure}

\begin{figure}[!ht]
\centering
\includegraphics[width=12cm,angle=0,keepaspectratio]{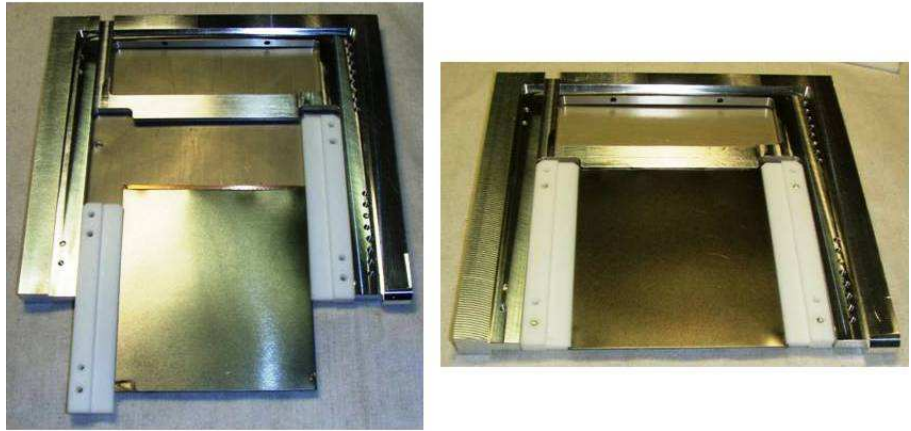}
\caption{\footnotesize The Aluminium holder and a blade (substrate
coated with $^{10}B_4C$ both sides) inserted.} \label{figMBv1prima3}
\end{figure}

\begin{figure}[!ht]
\centering
\includegraphics[width=8cm,angle=0,keepaspectratio]{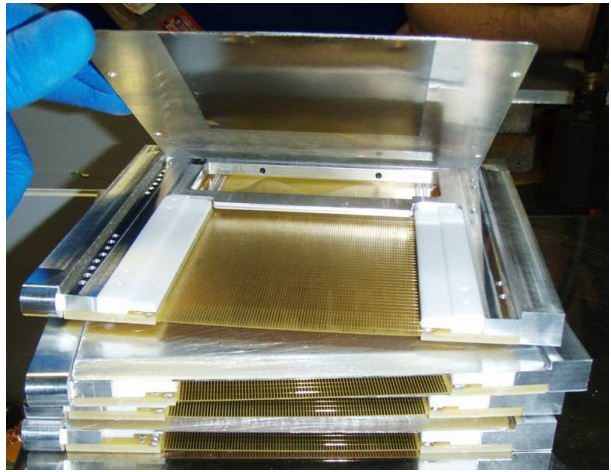}
\caption{\footnotesize Four cassettes assembled. The read-out PCB is
installed on the Aluminium holder.} \label{figMBv1prima4}
\end{figure}

\begin{figure}[!ht]
\centering
\includegraphics[width=12cm,angle=0,keepaspectratio]{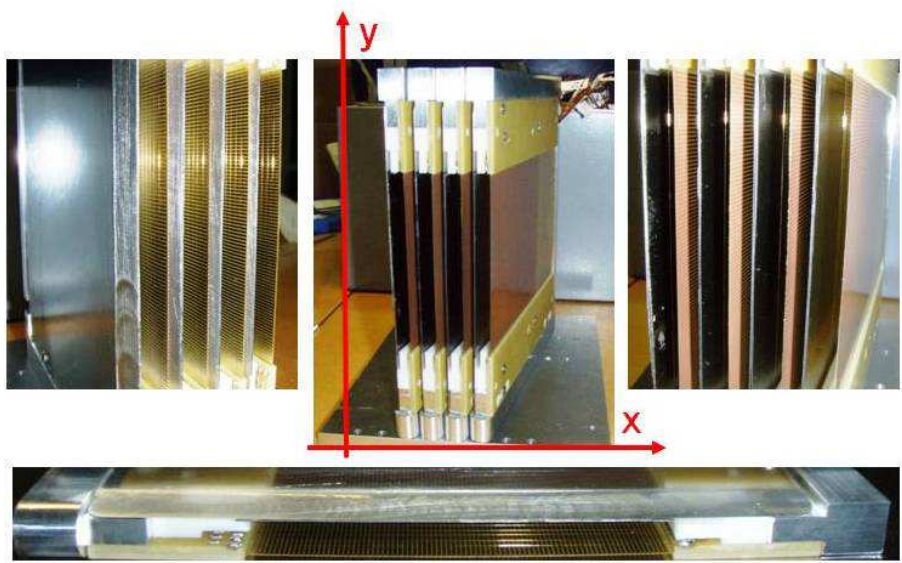}
\caption{\footnotesize Four cassettes stacked one after the other at
$10\,^{\circ}$ and parallel each other.} \label{figMBv1prima6}
\end{figure}

\begin{figure}[!ht]
\centering
\includegraphics[width=8cm,angle=0,keepaspectratio]{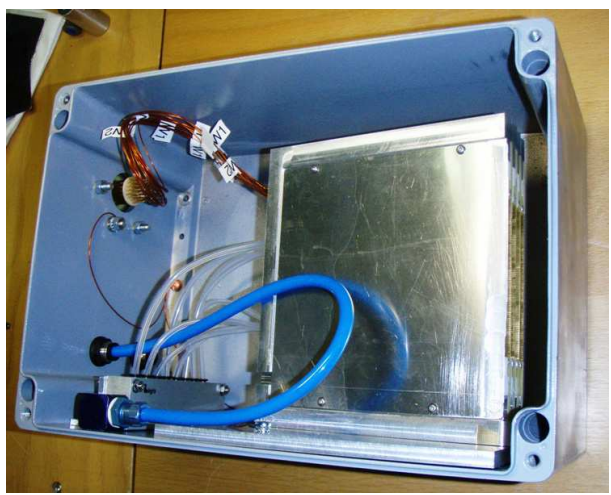}
\caption{\footnotesize The four cassette assembly in the gas vessel.
Each cassette is supplied by two gas inlets. The detector entrance
window is the one on the right.} \label{figMBv1prima7}
\end{figure}

\begin{figure}[!ht]
\centering
\includegraphics[width=10cm,angle=0,keepaspectratio]{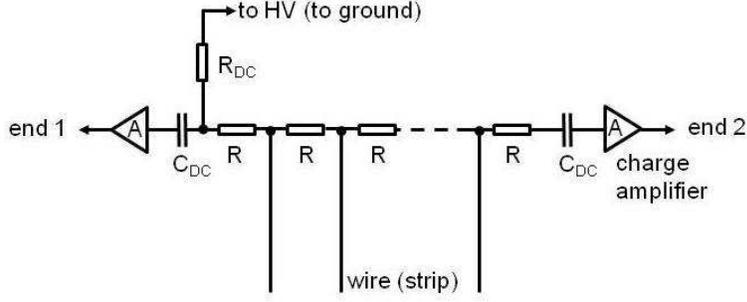}
\caption{\footnotesize The Multi-Blade front-end electronics
schematic.} \label{figMBv1prima8}
\end{figure}

The first prototype (V1) consists of four cassettes operated at
$10\,^{\circ}$. Given the cassette active region size, the prototype
active area, considering their overlap is about $6\times 9\,cm^2$.
\\ Figure \ref{figMBv1prima1} shows a polyimide PCB. The latter is
composed by a stack of three layers: two thick PCBs where in the
middle is fixed a $25\, \mu m$ polyimide foil. The inner part is
soft and the external part serves as a holder. 86 copper strips are
deposited on the surface of the thin region (see Figure
\ref{figMBv1prima2}).
\\ 39 anode wires (37 active wires and 2 guard wires)
are mounted and soldered on pads at $2\,mm$ distance from the
cathode plane. Both for anodes and for cathodes a resistive chain is
soldered on the rigid PCB. The total resistance is $6\,K\Omega$ for
the anode chain and $8\,K\Omega$ for the cathode chain. At the wire
plane edge a guard wire was installed to compensate the electric
field distortion, hence this wire will not produce any signal.
\\ The total gap between the converter and the cathode plane, i.e.
half cassette width, will be about $4\,mm$, thus any deformation of
either the substrate or the strip plane will produce a variation in
the local electric field produced between the anode plane and
cathodes. Consequently where the cathode is closer to the wire plane
the detector will manifest a higher gain. This effect mainly
degrades the uniformity over the cassette surface. The overall
uniformity on the whole detector surface is then degraded by the
single cassette uniformity and their arrangement in the space:
overlap and switching from one to another.
\\ It is crucial to control the flatness of both the substrate and
the PCB. The first manufactured PCB was composed of a thin polyimide
held on three sides by the rigid PCB. The provider was not able to
assure the polyimide flatness with this design. We changed the PCB
design in order to be able to pull on both sides and restore its
flatness. The polyimide is held by only two of its sides (see Figure
\ref{figMBv1prima1}). The PCB is held by a tool (see Figure
\ref{figMBv1prima1}) that allows to stretch the foil before being
mounted on the Aluminium holder. This tool allows also to mount
wires on the PCB keeping the system under tension. The $15\, \mu m$
tungsten wires are mounted on the PCB under a tension of $35\,g$.
Once wiring is over, the PCB can be installed on the holder.
\\ Figure \ref{figMBv1prima3} shows the Aluminium holder where the double side coated substrate with $^{10}B_4C$
\cite{carina} is inserted. In order to keep the wire tension, the
PCB, without removing the stretching tool, can be placed on the
holder. The holder and the PCBs present four different fixation
screw shifted by $0.25\,mm$ from each other. The PCB can be screwed
on the holder according to its actual size after stretching. This
ensures the right tension on the wires and the flatness of the
cathode plane.
\\ As for the read-out plane, the converter holding substrate must
be flat too. After sputtering, between the $Al$-substrate and the
$^{10}B_4C$ coating, a significant residual stress remains due to
the difference in the thermal expansion coefficient of $Al$
($\sim23.5\cdot10^{-6}\,1/K$) and $^{10}B_4C$
($\sim5.6\cdot10^{-6}\,1/K$). When they are cooled down to room
temperature the $Al$ contracts more than $^{10}B_4C$. Experiential
evidences of that have been observed: on single side coated
substrates, the un-coated side is shorter than the coated side,
resulting into a bending of the blade. When a double-side coated
blade has to be inserted into the holder (see Figure
\ref{figMBv1prima3}) the constraints on the sides make the blade
bend and unstable. The gain in efficiency of using a double layer
(option B) introduces some mechanical constraints which make its
realization much more complicated than the implementation of the
option A. We can deal with a reduced efficiency gaining in a
mechanical simplicity.
\\ We wanted to study both option A and B with this prototype but due
to the blade mechanical issue we convert the prototype in a single
layer detector.
\\ In order to keep the substrate with the converter flat enough to ensure a uniform
electric field, we mounted it on an Aluminium lid placed where a PCB
was removed (see Figure \ref{figMBv1prima4}). We used a $3\,\mu m$
thickness $^{10}B_4C$ coating instead. The gap between the wires and
the converter was increased up to $6\,mm$, while the gap between the
wires and the strips is about $2\,mm$. The MWPC is asymmetric.
Figure \ref{figMBv1prima4} shows the four cassettes equipped with
the read-out systems and the converters.
\\ The version V1 of the Multi-Blade detector allowed only to study
the single layer configuration A, as mechanical issues had made
impossible the initial configuration B realization.
\\ The number of read-out channels per cassette were reduced from 8 to 4: 2 anode and 2 cathode outputs.
\\ Four cassettes were stacked at $10\,^{\circ}$ with respect to the beam and parallel to each
other. Figure \ref{figMBv1prima6} shows the detail of the four
cassettes stacked from two points of view. We define as the
$x$-coordinate where the wire pitch is projected at $10\,^{\circ}$.
The $y$-coordinate is defined by the direction orthogonal to the strips orientation. \\
The four cassettes were then installed in the gas vessel, see Figure
\ref{figMBv1prima7}. Each cassette is supplied by two inlets to let
the gas flow directly inside them. The entrance window of the
detector is the one on the right in Figure \ref{figMBv1prima7}.
\\ The front-end electronics of the prototype is connected outside the gas vessel
and consists of a decoupling circuit and charge amplifiers. A
schematic of the whole front-end electronic chain is shown in Figure
\ref{figMBv1prima8}. Both wires and strips are connected in the same
way by their resistive chain, the AC signal is decoupled by two
capacitors at both ends from the DC component used to polarize the
wires at the HV and the strips to the ground potential.
\\ The charge is amplified by charge amplifiers. We used inverting
amplifiers of $6\,V/pC$ and $1\,\mu s$ shaping time for anodes and
non-inverting amplifiers of $32\,V/pC$ and $2\,\mu s$ shaping time
for cathodes. \\ Each chain ends into two signal outputs that can be
either summed to get the energy information (PHS) or subtracted and
divided to get the positional information.
\subsection{Results}
\subsubsection{Operational voltage}
\begin{figure}[!ht]
\centering
\includegraphics[width=7.5cm,angle=0,keepaspectratio]{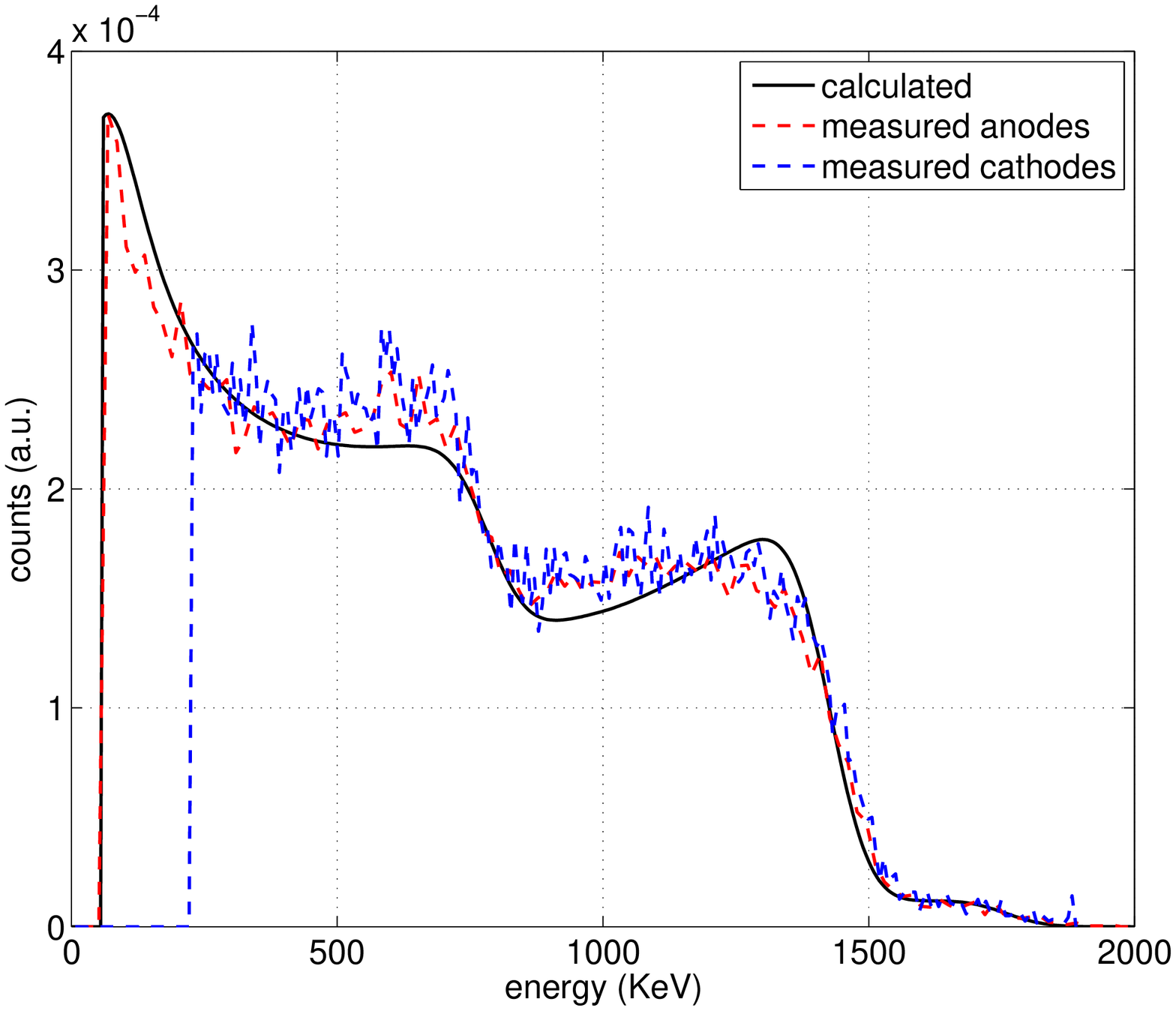}
\includegraphics[width=7.5cm,angle=0,keepaspectratio]{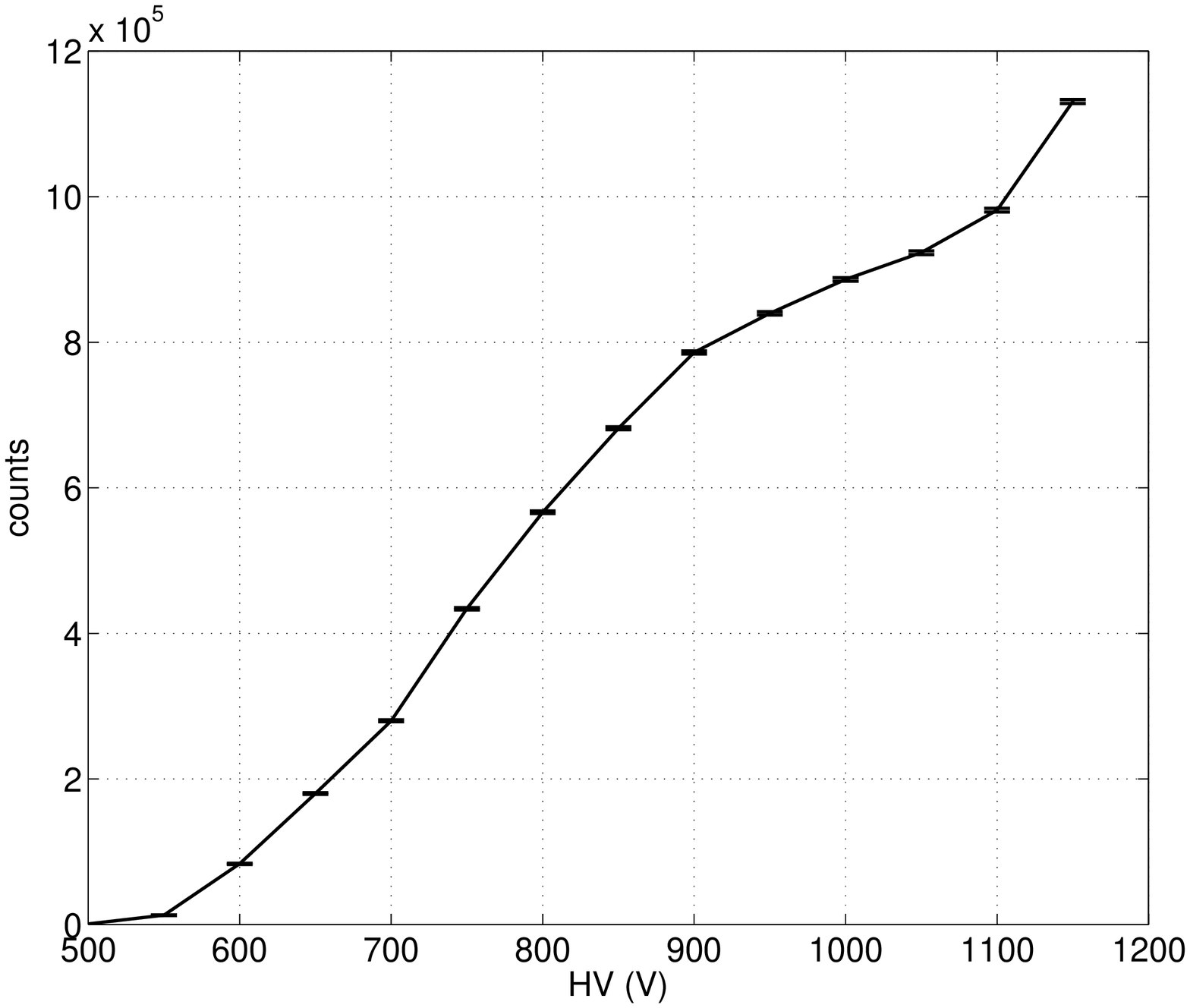}
\caption{\footnotesize PHS measured on strips and wires at $1000\,V$
and calculated PHS (left). The Multi-Blade detector plateau
(right).}\label{phspaltembv1456}
\end{figure}

A counting curve was measured in order to set the right bias voltage
to be applied to polarize the prototype. Each cassette output was
connected to get the energy information and then to measure the PHS.
Given the electronic noise, a $25\,mV$ threshold was used for the
anode amplifiers; for the cathode amplifiers we used $100\,mV$.
Figure \ref{phspaltembv1456} shows a PHS for both strips and wires,
compared with a PHS calculated according to \cite{fratheo} at
$1000\,V$.
\\The working voltage chosen based on the counting curve is $1000\,V$.
\subsubsection{Gain}
The Multi-Blade prototype is operated in proportional mode, its gain
has been measured on CT2 at ILL. The neutron flux of
$(15280\pm20)neutrons/s$ ($2.5$\AA) was quantified using the
procedure explained in details in \cite{jonisorma}.
\\ The prototype was polarized at $1000\,V$. Its $3\,\mu m$ $^{10}B_4C$ converter layer
was exposed to the beam orthogonally and
$\Phi_d=(904\pm2)neutrons/s$ were counted. It results into a
$\varepsilon=(5.92\pm4)\%$ detection efficiency. The current flowing
through the detector was measured and it is about
$I_{prop.}=180\,pA$.
\\ The detector operational voltage was set to $100\,V$.
The measurement of the current output was repeated operating the
detector in ionization mode, resulting into $I_{ion.}=3.1\,pA$.
\\ The average charge created for a detected neutron both proportional and ionization modes are:
\begin{equation}
Q_{prop.}=\frac{I_{prop.}}{\Phi_d}=199\,fC/neutron, \qquad
Q_{ion.}=\frac{I_{ion.}}{\Phi_d}=3.4\,fC/neutron
\end{equation}
This results into a gain of about $G=58$ at $1000\,V$.
\subsubsection{Efficiency}
Detection efficiency of the Multi-Blade prototype V1 has been
measured on CT2 at ILL by using a collimated and calibrated neutron
beam of wavelength $2.5$\AA.
\\ The neutron beam was calibrated using an $^3He$-based detector as explained in \cite{jonisorma}.
After the calibration, the neutron flux the Multi-Blade was exposed
to is $\left(15280\pm20\right)\,neutrons/s$ over an area of $2\times
7\,mm^2$.
\\ The efficiency was measured for the following bias voltages
$950\,V$, $1000\,V$ and $1050\,V$. The efficiency was measured on
the four cassettes under the angle of $10^{\circ}$ and then
averaged. The results are listed in Table \ref{tabdeghjknflta99}.
\begin{table}[!ht]
\caption{\footnotesize Multi-Blade detection efficiency at $2.5$\AA
\, for three bias voltages and calculated efficiency \cite{fratheo}
for a given threshold.} \label{tabdeghjknflta99} \centering
\begin{tabular}{|c|c|c|c|}
\hline \hline $HV (V)$ &  $\varepsilon $(at$ \,\, 2.5$\AA) & Threshold $(KeV)$ & calculated $\varepsilon $(at$ \,\, 2.5$ \AA)\\
\hline
$950$     &  $\left( 24.4 \pm 0.2\right)\%$  & $100$ & $25.7\%$  \\
$1000$    &  $\left( 25.4 \pm 0.2\right)\%$  & $70$  & $26.7\%$ \\
$1050$    &  $\left( 26.0 \pm 0.2\right)\%$  & $50$  & $27.2\%$  \\
\hline \hline
\end{tabular}
\end{table}
\\ The result is in a good agreement with what can be calculated
from the theory in \cite{gregor} and \cite{fratheo} by using an
energy threshold of $100\,KeV$, $70\,KeV$ and $50\,KeV$ for the
three voltages from $950\,V$ to $1050\,V$ respectively. These
thresholds have been determined from the comparison between the
measured and calculated PHS \cite{fratheo}, as shown in Figure
\ref{phspaltembv1456} for the bias voltage of $1000\,V$.
\subsubsection{Uniformity} 
\begin{figure}[!ht]
\centering
\includegraphics[width=7.5cm,angle=0,keepaspectratio]{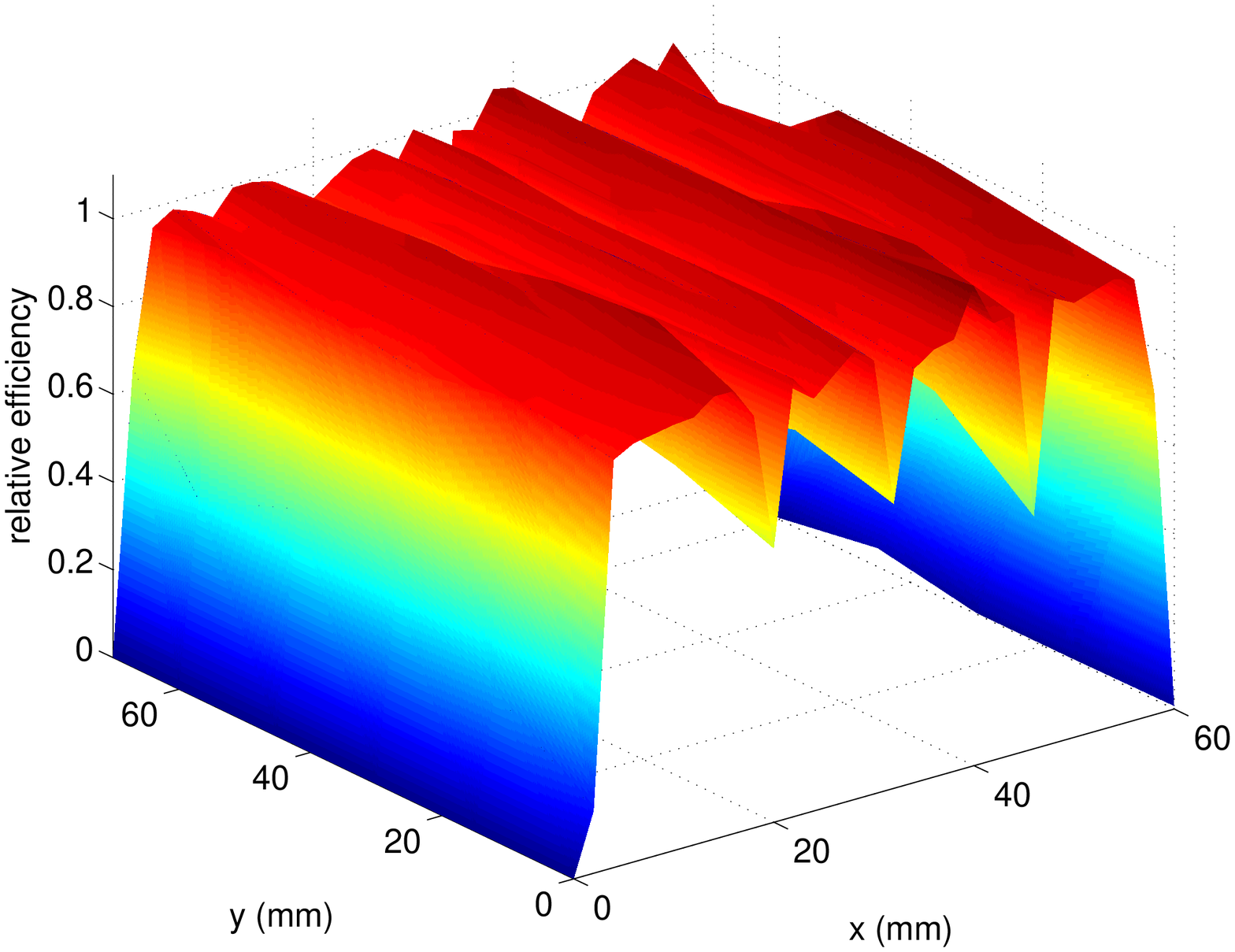}
\includegraphics[width=7.5cm,angle=0,keepaspectratio]{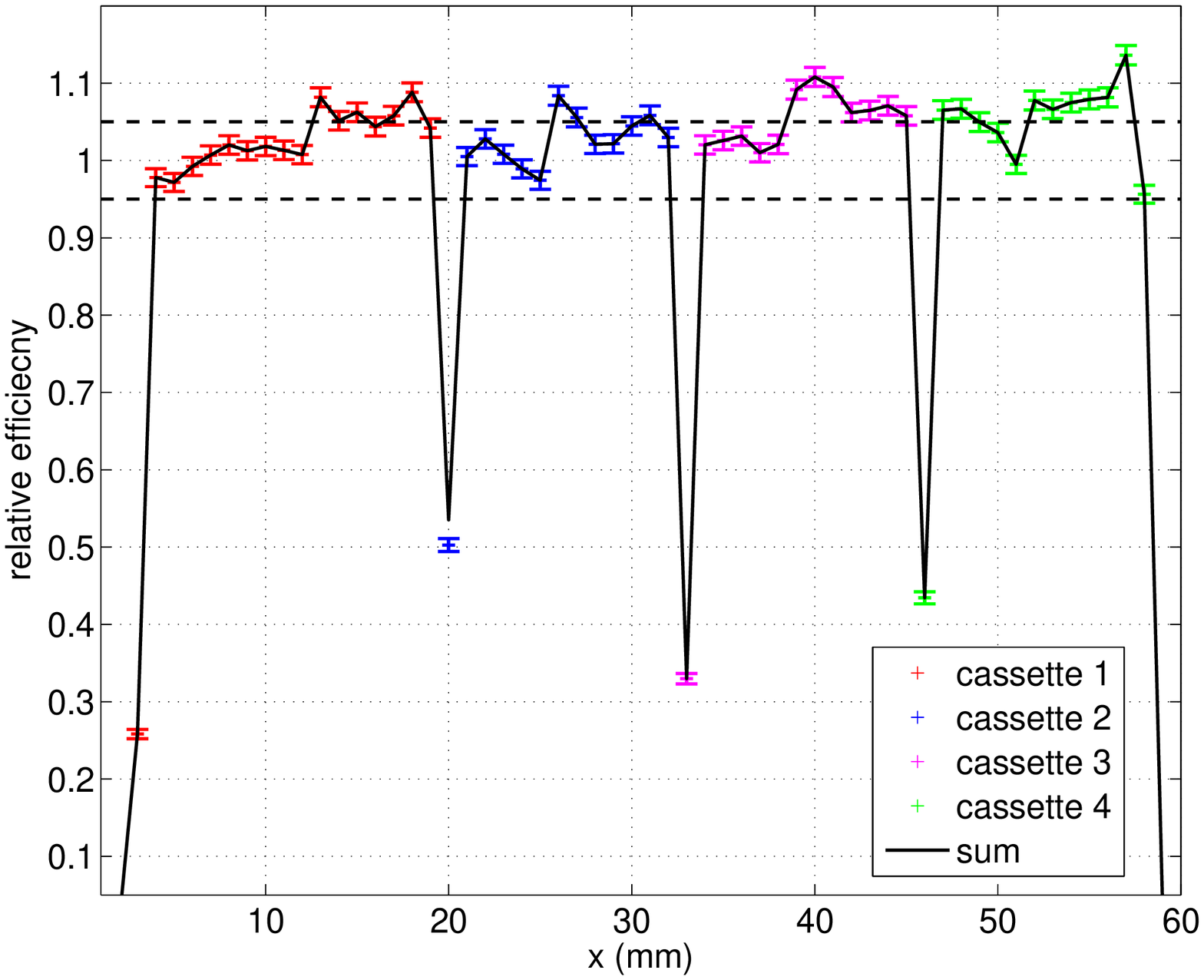}
\caption{\footnotesize Relative efficiency scan over the whole
detector surface.}\label{crgw5465u6u}
\end{figure}

The main issue in the Multi-Blade
design is the uniformity over its active surface. The cassettes
overlap to avoid dead zones, and, in the switching between one
cassette to another, a loss in efficiency can occur. There are
mainly two reasons that cause the efficiency drop: at the cassette
edge the electric field is not uniform and there is some material
that scatters neutron on the way to the next cassette. In order to
reduce the dead zone at the cassette edge, i.e. any material that
can cause scattering, each cassette is cropped (see Figure
\ref{figMBv1prima6}) to be parallel to the incoming neutron
direction.
\\ Moreover, when a neutron is converted at the cassette edge, it
produces a fragment that half of the time travels toward outside of
the cassette and half time inward. We expect not to have generated
charge for about $50\%$ of the events. In addition to that the
electric field at the edge may not be uniform and the guard wire
contributes to enlarge the dead zone because it does not generate
charge amplification.
\\ We scan with a collimated neutron beam and we register a PHS over
the whole prototype surface by using a $1\,mm$ step for the
$x$-direction and a $10\,mm$ step for the $y$-direction. We
integrate the PHS for each position and we obtain a local counting.
We normalize it to 1 on the average efficiency. Figure
\ref{crgw5465u6u} shows the relative efficiency scan over the whole
detector. The scan along the cassettes in the position $y=40\,mm$ is
also shown.
\\ Each cassette shows a quite uniform response
along its strips ($y$-direction); the maximum efficiency relative
variation is below $2\%$. On the other hand, in the gap between two
cassettes the efficiency drops about $50\%$ in a region which is
$2\,mm$ wide.
\subsubsection{Spatial resolution}
\begin{figure}[!ht]
\centering
\includegraphics[width=7.5cm,angle=0,keepaspectratio]{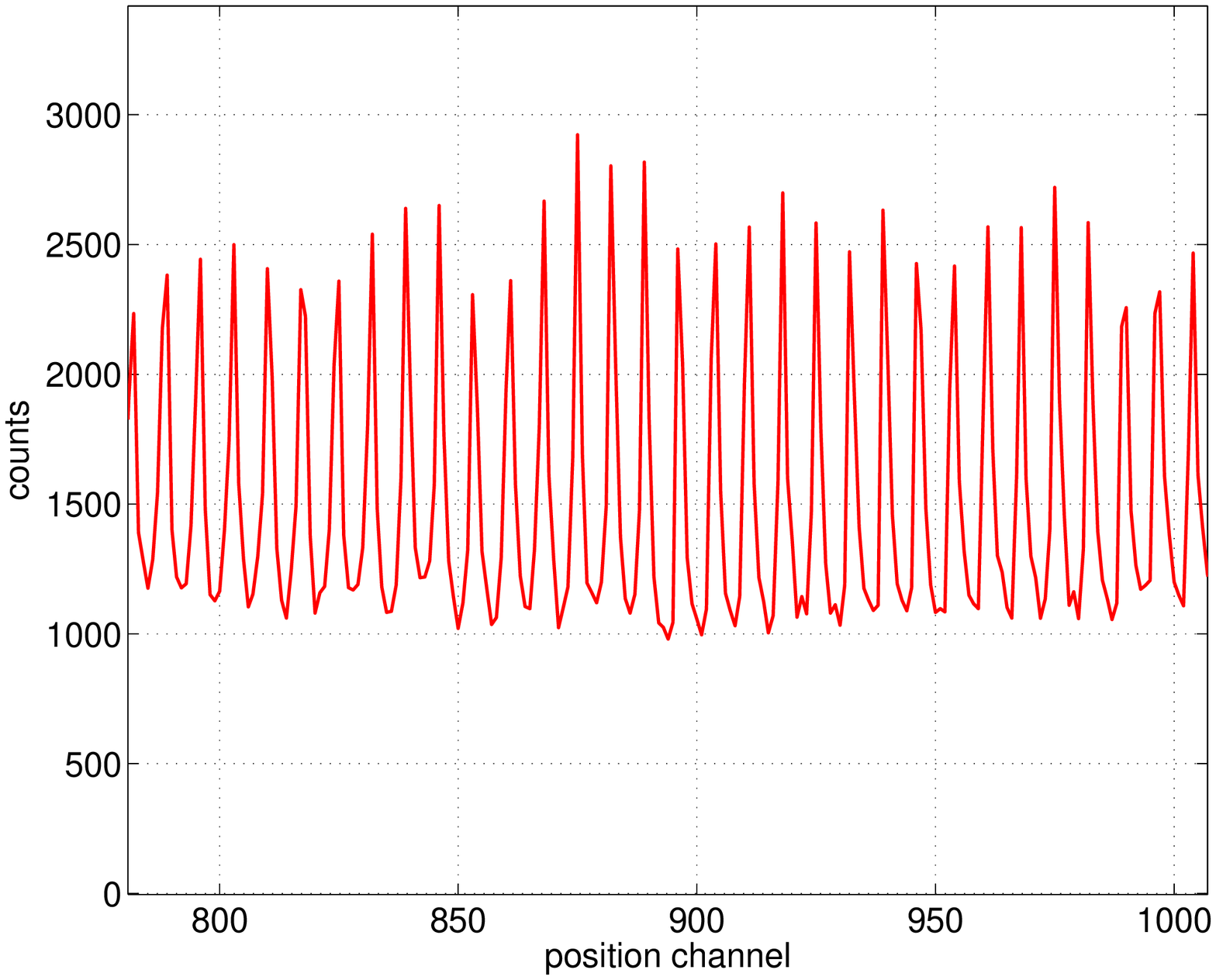}
\includegraphics[width=7.5cm,angle=0,keepaspectratio]{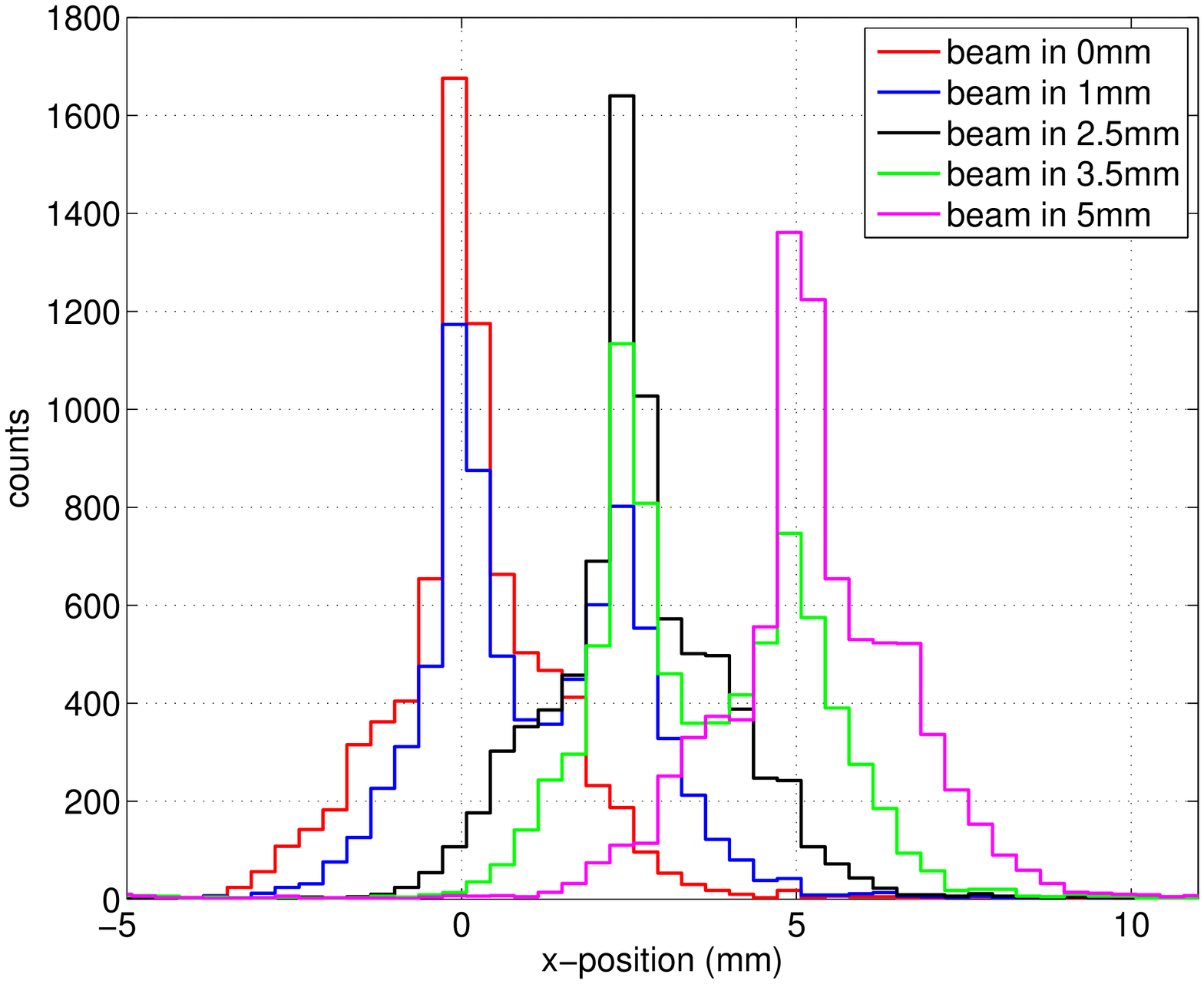}
\caption{\footnotesize Diffuse beam wire response in charge division
(left), collimated neutron beam response as a function of the beam
position (right).}\label{figurewirerspsv1}
\end{figure}

\begin{figure}[!ht]
\centering
\includegraphics[width=10cm,angle=0,keepaspectratio]{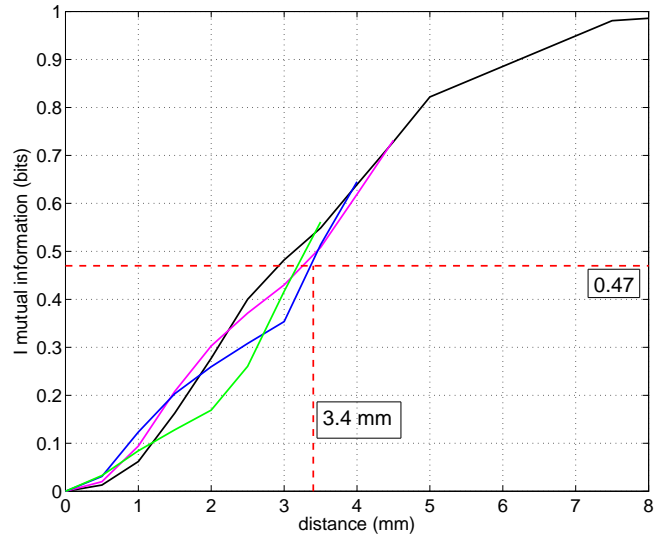}
\caption{\footnotesize Mutual information as a function of the
distance between the response distributions of the neutron detector.
The horizontal line defines an information of $0.47$ bits that
corresponds to a $3.4\,mm$ spatial resolution (before projection) in
the worse case.}\label{spatfigu57}
\end{figure}

\begin{figure}[!ht]
\centering
\includegraphics[width=10cm,angle=0,keepaspectratio]{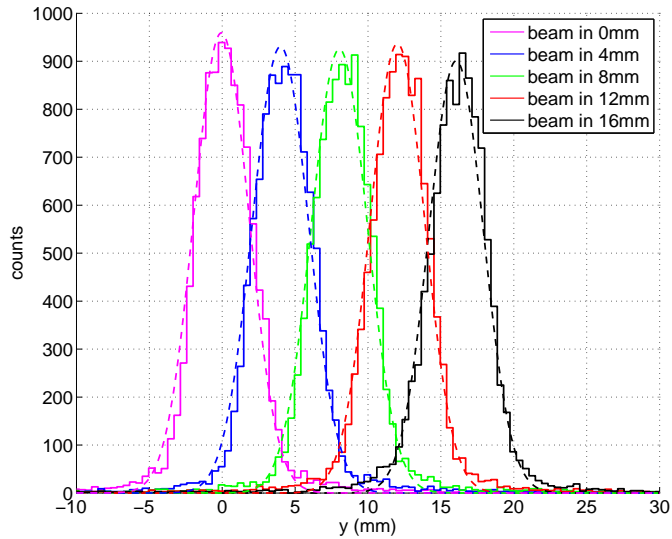}
\caption{\footnotesize Fine beam neutron scan along the strip
cathodes. The spatial resolution is given by the FWHM and
corresponds to $4.4\,mm$.}\label{figrespyresolmb4}
\end{figure}

In order to calculate a detector spatial resolution one should be
careful as to which definition has to be adopted in order to give a
meaningful result. If the detector response is a continuous function
or discrete the problem should be tackled in a different way.
\\ Here we refer to \cite{patrickinforis} to properly calculate the
spatial resolution given by the anodes (discrete response function)
and to a standard Full Width Half Maximum (FWHM) definition for what
concerns the cathodes (continuous response function) of the
Multi-Blade prototype.
\paragraph{Spatial resolution: x}
The version V1 of the Multi-Blade prototype is operated at
$10^{\circ}$ between the neutron incoming direction and the detector
converter layer. We recall that the wire plane is projected on the
neutron incoming direction. An improvement by a factor
$\sin(10^{\circ})\sim0.17$ is achieved on the horizontal spatial
resolution with respect to an orthogonal incidence. E.g. if the
spatial resolution, before projection, were a wire pitch (in our
prototype $2.5\,mm$) this results in an actual resolution of about
$0.45\,mm$.
\\ In Figure \ref{figurewirerspsv1} the charge division
response of the wire plane is shown by using either a diffuse beam
or a collimated beam down to $1\,mm$ footprint ($2.5$\AA).
\\ The charge division method is able to identify each wire anode
position; when a collimated beam is in one position we can either
get a single wire or two wires reacting. This effect is due to the
fact that the wire plane splits the gas volume into almost
independent cells, thus the charge generated by primary ionization
in one cell makes its associated wire react. In our case, since we
are using a mixture of $Ar/Co_2$ ($90/10$) at atmospheric pressure
the $^{10}B$ neutron capture reaction fragment ranges make a few
$mm$. Therefore, if the track is contained in one single wire cell,
only a single wire reacts; on the other hand if the track travels
across two cells we get a two wire response. Since the wire plane is
read-out in charge division, if two wires react, the hit will be
identified to be in between the two wires, corresponding to the
charge centroid. The response distribution, for a given beam
position, will have tails corresponding to these events. Figure
\ref{figurewirerspsv1} shows the resulting distribution as the
neutron beam moves along the detector. One can wonder now what is
the actual spatial resolution in this situation. In order to
quantify it is necessary to apply the informational-theoretical
approach of \cite{patrickinforis}.
\\ We calculate the mutual information between the distribution in
Figure \ref{figurewirerspsv1} for all the possible combinations and
we will take as resolution the worse result at an information
threshold level of $0.47$ bits which corresponds to the FWHM
criterion of a continuous distribution.
\\ Figure \ref{spatfigu57} shows the mutual information as a
function of the distance of the neutron distribution response of our
detector for different reference positions \cite{patrickinforis}. We
notice that in the worst case we end up with $3.4\,mm$; which
translates into a spatial resolution of $0.6\,mm$ at $10^{\circ}$.
\\Note that the spatial resolution lies in between
a single wire pitch ($2.5\,mm$) and two.
\paragraph{Spatial resolution: y}
As already mentioned, particle tracks in our gas make a few $mm$.
Since the cathodes read-out strips are $0.8\,mm$ wide and they are
spaced by $0.2\,mm$ and the read-out is performed by a charge
division chain, there are several strips that are involved in the
induction process per event. The charge division determines the
charge centroid along the $y$-direction in the detector.
\\ For the cathodes the response can be considered continuous and the
FWHM method is suitable.
\\ Figure \ref{figrespyresolmb4} shows the strip response as a
function of the position of the collimated beam hitting the
detector. By performing a gaussian fit we obtain a spatial
resolution (FWHM) for the vertical direction $y$ of about $4.4\,mm$.

\section{Multi-Blade version V2}\label{sectv2mbg}
\subsection{Mechanical study}
\begin{figure}[!ht]
\centering
\includegraphics[width=4.5cm,angle=0,keepaspectratio]{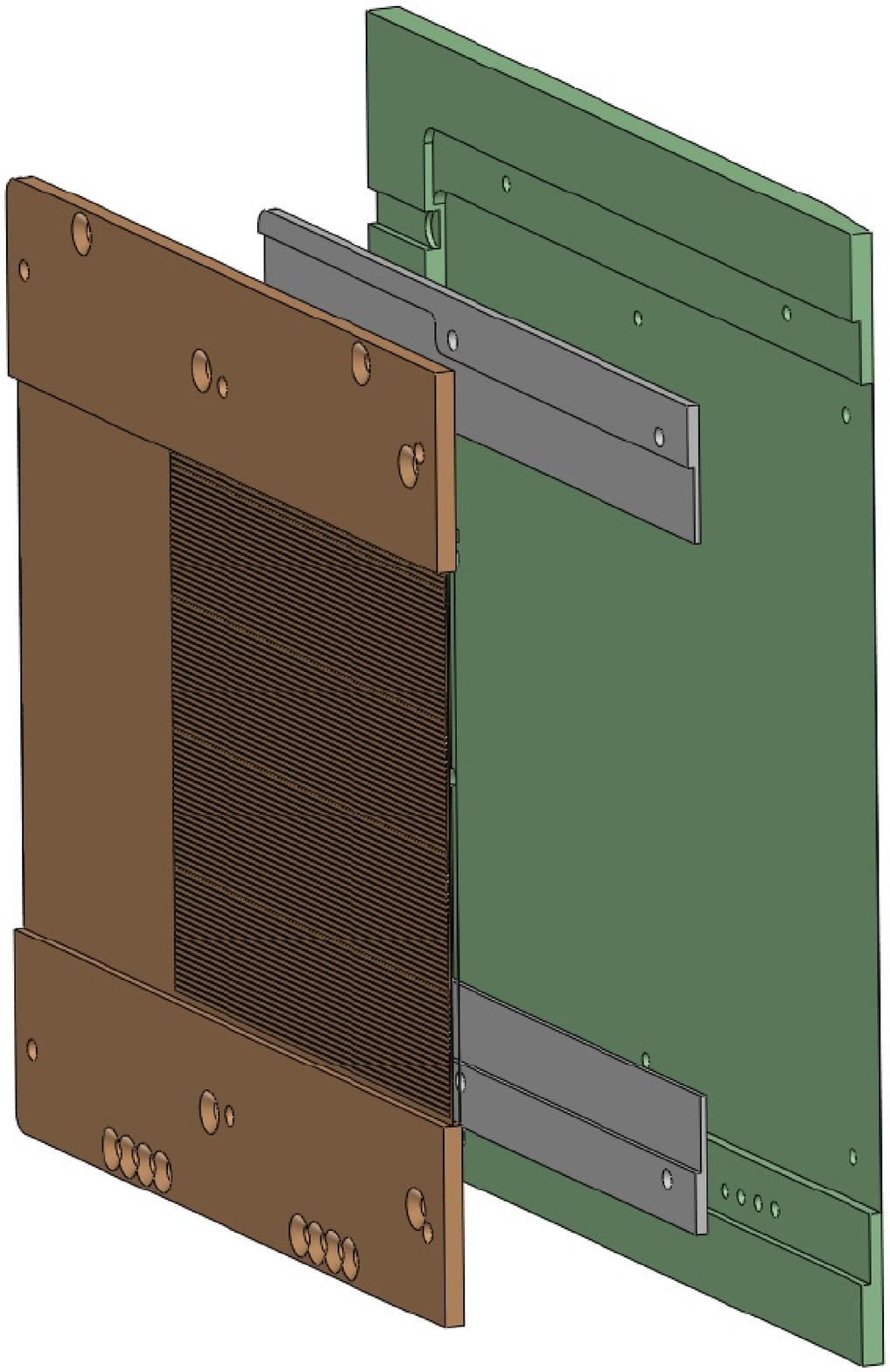}
\includegraphics[width=8.2cm,angle=0,keepaspectratio]{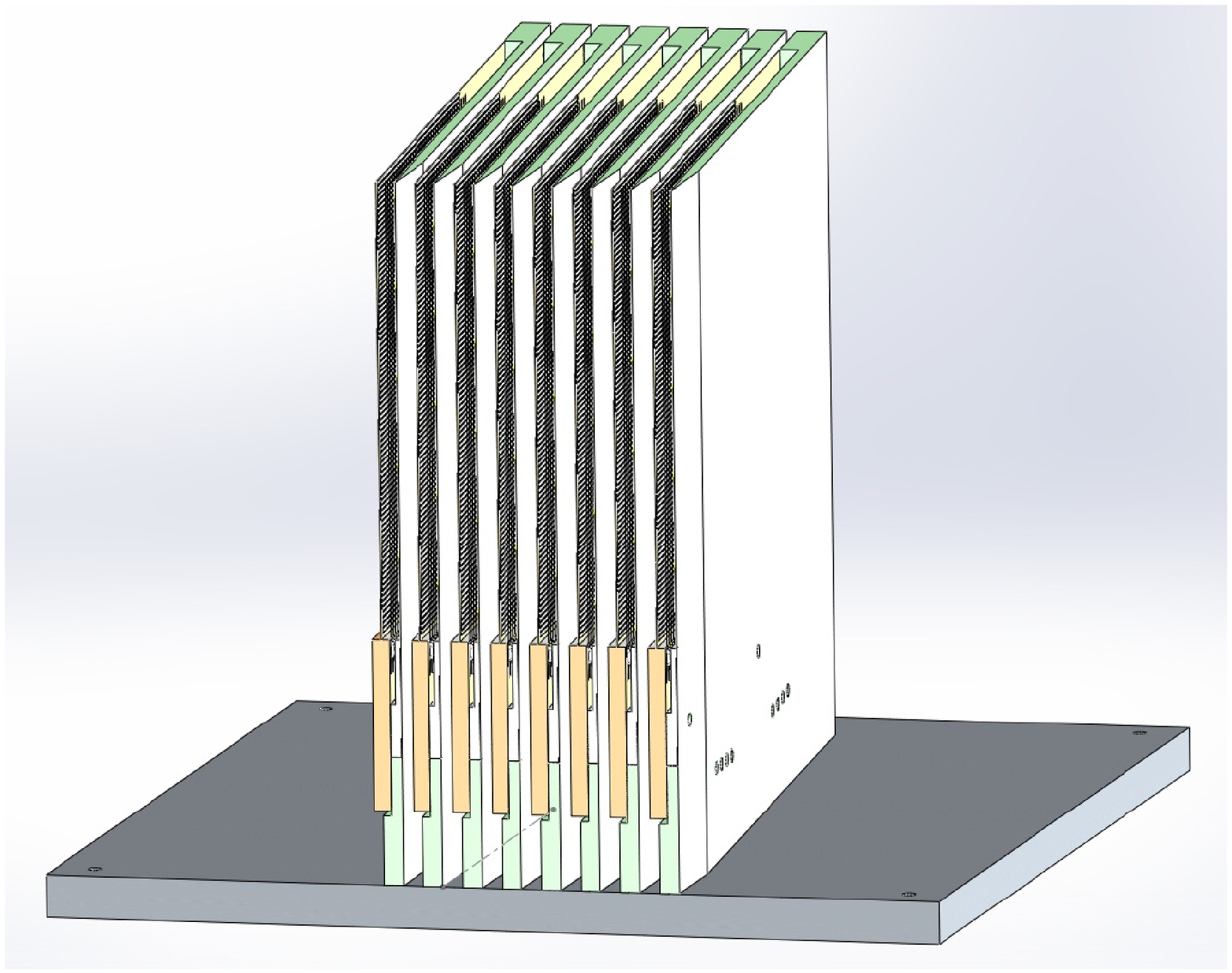}
\caption{\footnotesize A cassette conceived to hold one converter
layer (left) and a stack of several cassette (right).}
\label{mbv2scehm543ergg}
\end{figure}

We learned from the Multi-Blade version V1 that the single layer
configuration (option A) presents less mechanical constraints.
Moreover, the substrate holding the converter has not to be crossed
by neutrons that makes its manufacture easier.
\\ The converter layer can be thick because the efficiency is
saturated above $3\,\mu m$; the substrate can be thick also because
it has not to be crossed by neutrons to hit a second converter.
Hence, the substrate can be an integrated part of the cassette
holder, the converter layer can be directly deposited over its
surface. The read-out system used is the same as in version V1.
Neutrons have still to cross the PCBs before being converted. Figure
\ref{mbv2scehm543ergg} shows a cassette and a stack of them
conceived for the single layer option.
\\ The cassettes are oriented at $5\,^{\circ}$ with respect to the incoming neutron
direction. The sensitive area of each cassette is $10\times 9\,cm^2$
but, the actual sensitive area offered to the sample is given by
$(10\,cm \cdot \sin(5\,^{\circ}))\times 9\,cm = 0.9\times 9\, cm^2$.
As a result, the actual projected wire pitch is improved down to
$0.22\,mm$.
\subsection{Mechanics}
\begin{figure}[!ht]
\centering
\includegraphics[width=12cm,angle=0,keepaspectratio]{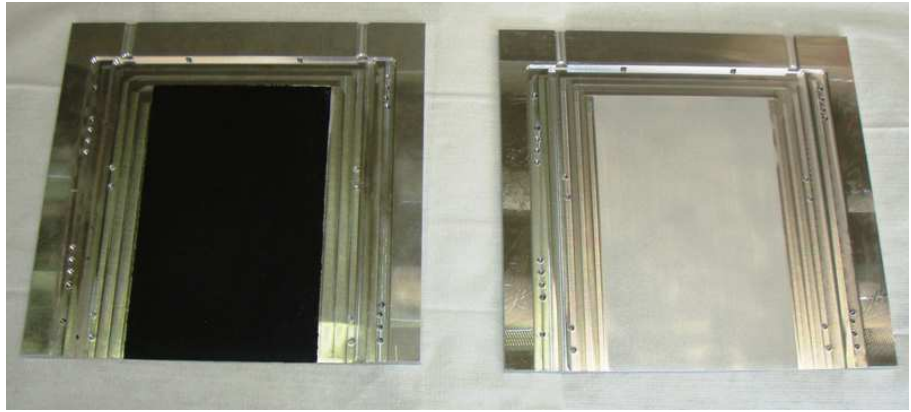}
\caption{\footnotesize A cassette V2 coated and un-coated with
$^{10}B$ painting.} \label{asd56}
\end{figure}

\begin{figure}[!ht]
\centering
\includegraphics[width=10cm,angle=0,keepaspectratio]{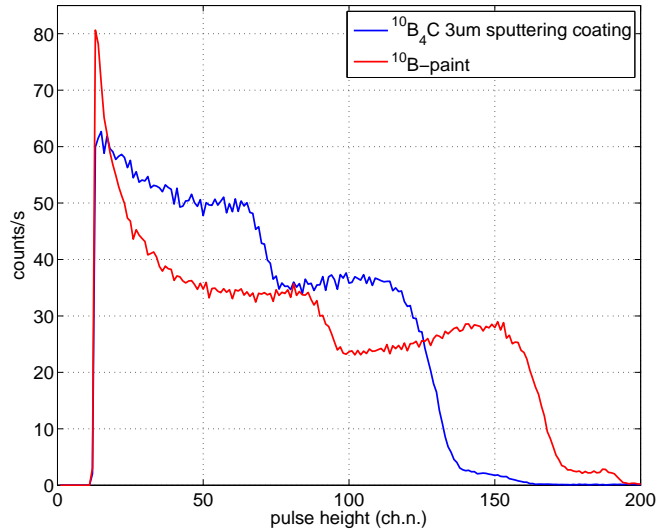}
\caption{\footnotesize Comparison between the PHS of the sputtered
coated layers \cite{carina} and the $^{10}B$ painting.}
\label{eretvrtr57457}
\end{figure}

\begin{figure}[!ht]
\centering
\includegraphics[width=8cm,angle=0,keepaspectratio]{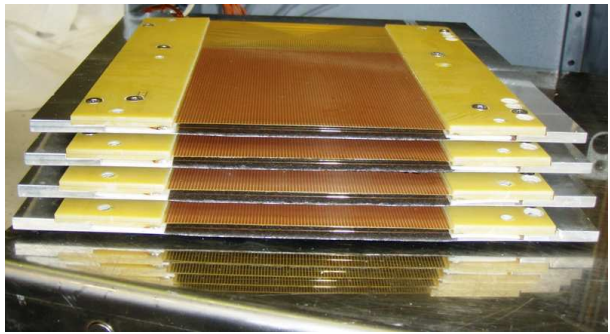}
\caption{\footnotesize Four fully assembled cassettes for the
Multi-Blade version V2.} \label{fulcassv2546}
\end{figure}

\begin{figure}[!ht]
\centering
\includegraphics[width=7cm,angle=0,keepaspectratio]{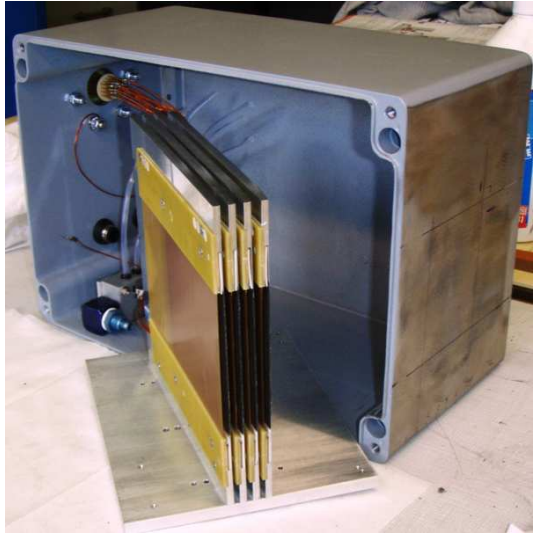}
\caption{\footnotesize The four cassettes installed at $5\,^{\circ}$
with respect to the detector window being installed in the gas
vessel.} \label{fullv246tf}
\end{figure}

The second prototype (V2) consists also of four cassettes but we
operate them at $5\,^{\circ}$. At this inclination the expected
efficiency at $2.5$\AA \, is about $43\%$ if we employ the sputtered
coating of the version V1 \cite{carina}. The cassettes are the ones
shown in Figure \ref{mbv2scehm543ergg}, conceived to study the
single converter layer option. A rigid substrate is directly coated
with the converter material. The cassette width in the version V1
was about $12\,mm$, in the version V2 we reduce their actual size to
$6\,mm$. Consequently the MWPC gap, between the converter and the
cathodes, is $4\,mm$. With respect to the version V1 the wire plane
is closer to the converter.
\\The prototype active area, considering the cassette overlap, is
about $3.2\times 9\,cm^2$.
\\ The read-out PCBs are those used in the version V1.
\\ Figure \ref{asd56} shows a cassette substrate both coated and
un-coated.
\\ Since the efficiency is saturated as the thickness of the layer
exceeds $3\,\mu m$, we study the possibility to use different
converters. We can deposit a painting containing $^{10}B$ grains and
make a coating a few hundreds of microns thick. A thick layer also
functions as an integrated collimator. Any neutron that comes from
the sides of the detector has less probability to be detected and is
more likely absorbed in the outer layers. Hence, only neutrons which
impinge the detector from the front have a serious chance to
generate a signal. Neutron background is then decreased.
\\ The uniformity of the coating, even in the single layer configuration, is an
important aspect to guarantee the converter flatness. The latter has
to ensure the precision of the neutron incidence angle, in fact if
it varies slightly the efficiency changes widely. Furthermore, a
deviation from the converter flatness also induces the variation of
the electric field and then the local gain of the detector changes.
\\ The roughness of the converter should be below the
neutron capture fragment ranges, which is of the order of a few $\mu
m$ for $^{10}B$. In fact, the gain in efficiency due to an
inclination comes from the fact that the neutron path travels close
under the surface. If the surface is irregular (on the $\mu m$ scale
or more), that can be seen as equivalent for a neutron to hit a
surface perpendicularly, there is not much gain in efficiency. It is
crucial that the size of our grains, in the painting, is less than
the particles ranges, i.e. their size should be below the micron
scale for $^{10}B$.
\\ The conductivity of the painting can be an issue.
If the resistivity is too large the charge evacuation is not
guaranteed and consequently the actual electric field is affected.
We mix a glue with $^{10}B$ grains of sizes $\sim10 \mu m$. We did
not have access to a finer-grained $^{10}B$ powder: our grinding
technique resulted in $\sim10\,\mu m$ grain size. We used this
powder. As the grain size is not smaller than the fragment ranges we
know that there can be an efficiency issue. The layer resistivity
was measured to be about $50\, M\Omega \cdot m$ and for a $0.5\,mm$
thick layer.
\\ Figure \ref{eretvrtr57457} shows two PHS: one is taken with a $3\,\mu m$ thick $^{10}B_4C$ layer \cite{carina}
and the other with the $^{10}B$ painting both installed in a MWPC,
at normal incidence and with a $2.5$\AA \, neutron beam. The
difference in gain on the two spectra is due to the difference in
the gas gap between the wire plane and the converter, since the
painting is a few $mm$ closer to the wires than the sputtered layer.
\\ The painting efficiency is $1.5\%$ lower than the sputtered
coating.
\\ In order to investigate if the resistivity of the painting is not to large to avoid the evacuation
of the charges, we place the painting layer on a very intense beam
of $560\,KHz$ and we measure the counting rate as a function of
time. There are no losses after several hours. The resistivity of
the $^{10}B$ painting seems to be acceptable.
\\ The painting is suitable for single layer application supposed that we can control the flatness
of the layer and to use smaller grains. We do not guarantee the
sputtered layers efficiency for our prototype under an angle because
of the size of the grains we used in the painting.
\\ The converter painting was not optimized, hence we expect some
problems due to its not perfect regularity. This effect will be more
evident at the edge of each cassette where the flatness affect to a
greater extent the electric field glitches.
\\ We mount the prototype using the painting. Four cassettes were assembled.
Figure \ref{fulcassv2546} and \ref{fullv246tf} show the cassettes
and the installation in the gas vessel for testing.
\\ The electronics used is the same as in the version V1 of the
Multi-Blade.
\subsection{Results}
\subsubsection{Operational voltage}
The measure of the counting curve gives the bias voltage of
$800\,V$.
\\ The operational voltage is lower than the one used for the first prototype
because the gap between anodes and cathodes was reduced in the new
design. For this reason the PHS is degraded also because the maximum
path in $Ar/CO_2$ of an $\alpha$-particle makes almost $9\,mm$ it is
more likely in the version V2 to hit the opposite cathode before
depositing its entire energy in the gas volume.
\begin{figure}[!ht]
\centering
\includegraphics[width=7.5cm,angle=0,keepaspectratio]{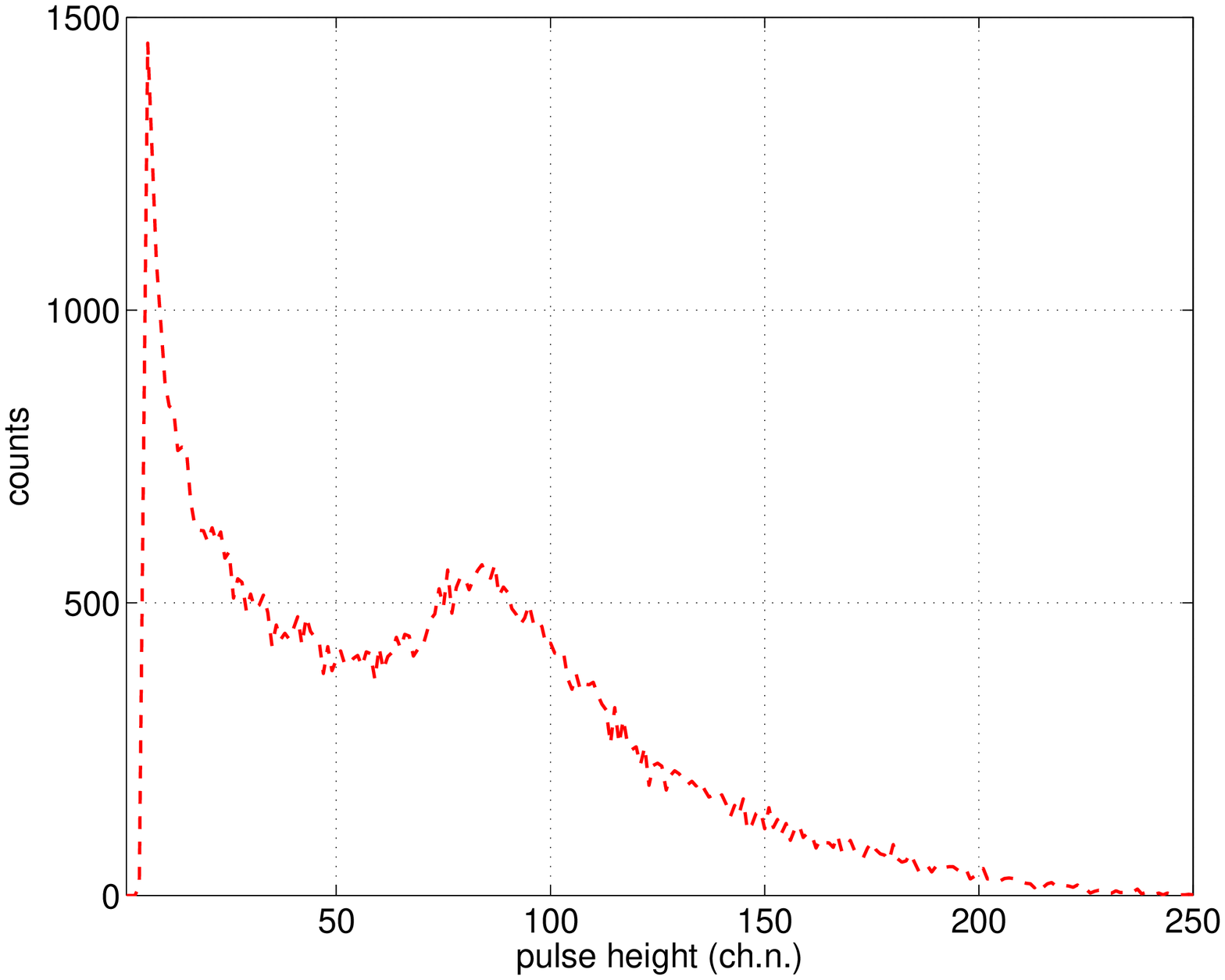}
\includegraphics[width=7.5cm,angle=0,keepaspectratio]{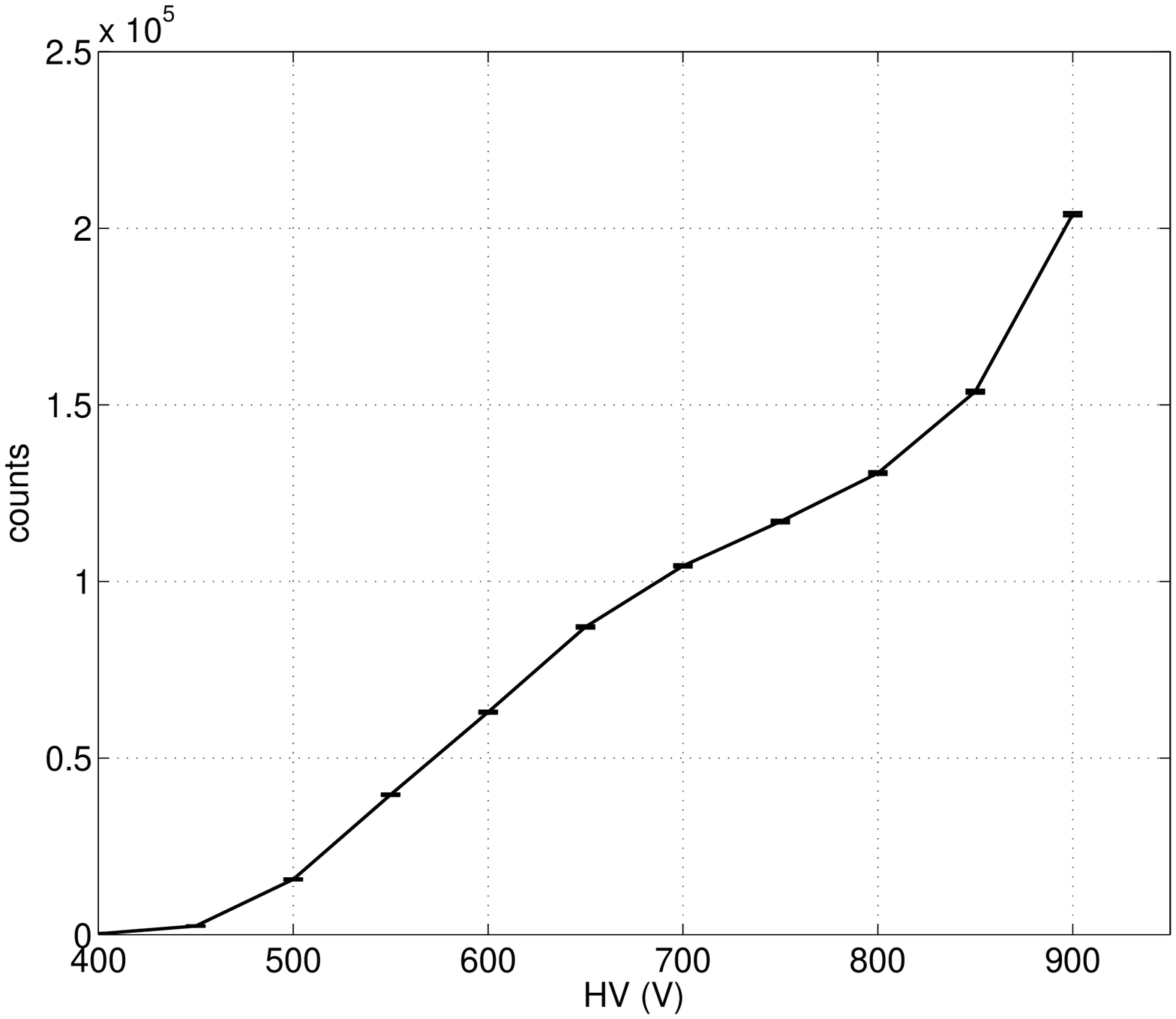}
\caption{\footnotesize PHS measured on wires at $800\,V$ (left). The
Multi-Blade detector plateau (right).}\label{phspaltembv1456ewc5}
\end{figure}
\subsubsection{Efficiency}
Detection efficiency of the Multi-Blade prototype V2 has been
measured on CT2 at ILL by using a collimated and calibrated neutron
beam of wavelength $2.5$\AA using the procedure explained in
\cite{jonisorma}.
\\ The neutron flux used is $\left(12010\pm20\right)\,neutrons/s$
over an area of $2\times 6\,mm^2$.
\\ The efficiency was measured for the operational voltage $800\,V$.
The efficiency was measured on the four cassettes under the angle of
$5^{\circ}$ and then averaged. The results is:
\begin{equation}
\varepsilon\left( \mbox{at }2.5\mbox{\AA}\right)=\left( 8.32 \pm
0.05\right)\%
\end{equation}
The expected efficiency for a sputtered layer of which the roughness
is widely below the $\mu m$ scale is about $43\%$ (at $2.5$\AA). In
our case, however the inclination only increases the efficiency by a
few percent because of the grains size. The surface irregularity
makes the inclination effect vanish. Neutrons only impinge almost
perpendicularly on the microscopic grain structure.
\\ By using a sputtered layer \cite{carina} or, if one can better
control the painting flatness and have smaller grain size, there
should be no reason not to get the calculated efficiency.
\subsubsection{Uniformity}
\begin{figure}[!ht]
\centering
\includegraphics[width=10cm,angle=0,keepaspectratio]{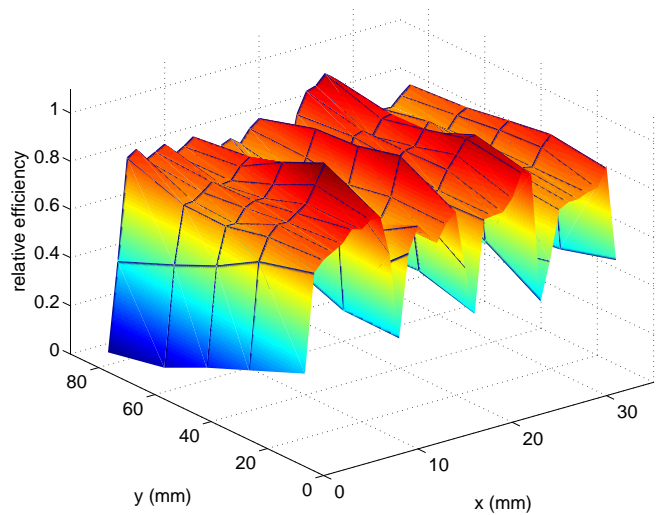}
\caption{\footnotesize Relative efficiency scan over the whole
detector surface.} \label{unifgasf945ntgww54}
\end{figure}

In the version V2 of the Multi-Blade the mechanics is more compact
in order to avoid dead zones in the overlap between the cassettes.
It has been discussed that the two issues which degrade the
uniformity are those dead zones and the electric field at each
cassette edge.
\\ Even though the mechanics design is more efficient in the version V2, the electric field
issue remains.
\\Moreover, the coating done by the painting was slightly irregular
mostly at the edges of each cassette. This diminishes the precision
by which we switch between one and the following. A large amount of
converter material at the edge will absorb most of the neutrons
without generating any signal.
\\ Figure \ref{unifgasf945ntgww54} shows the relative efficiency scan over the
detector surface. Compared with the version V1 uniformity is worse.
The gap between the converter and the wire plane is $2\,mm$, hence
any irregularity on the layer surface will affect the local gain of
the detector. Even along each cassette ($y$-direction) the gain
varies by about $10\%$ while in the sputtered version it only varied
of about $2\%$.
\\ Moreover, now looking at the $x$-direction, at the cassettes edge
the efficiency now drops more than $50\%$. This is due to the amount
of converter at the edge that does not generate signal but only
absorbs neutrons.
\\ The version V2 is in principle more compact and if the coating
would be precise the uniformity was expected to be better than in
the version V1.

\subsubsection{Spatial resolution}
\begin{figure}[!ht]
\centering
\includegraphics[width=12cm,angle=0,keepaspectratio]{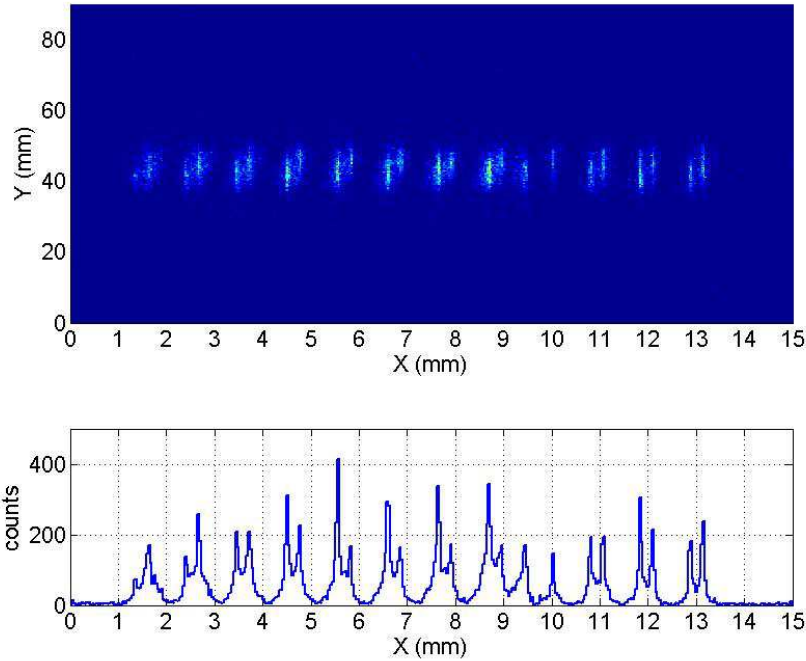}
\caption{\footnotesize An image and its projection on the $x$ axis
taken with the prototype. Each slit is $0.2\,mm\times10\,mm$ large
and it is spaced by $0.9\,mm$.} \label{figexp0n74jdfiwhud}
\end{figure}

\begin{figure}[!ht]
\centering
\includegraphics[width=10cm,angle=0,keepaspectratio]{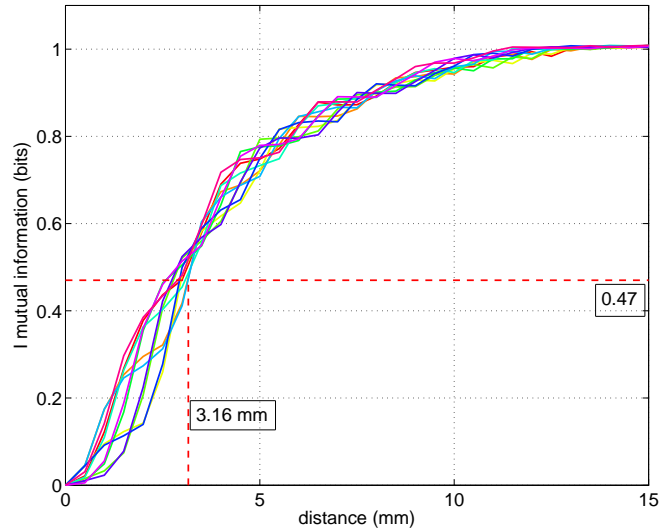}
\caption{\footnotesize Mutual information as a function of the
distance between the response distributions of the neutron detector.
The horizontal line defines an information of $0.47$ bits that
corresponds to a $3.16\,mm$ spatial resolution (before projection)
in the worse case.} \label{mutinforfgjk830nsbbcd}
\end{figure}

\begin{figure}[!ht]
\centering
\includegraphics[width=10cm,angle=0,keepaspectratio]{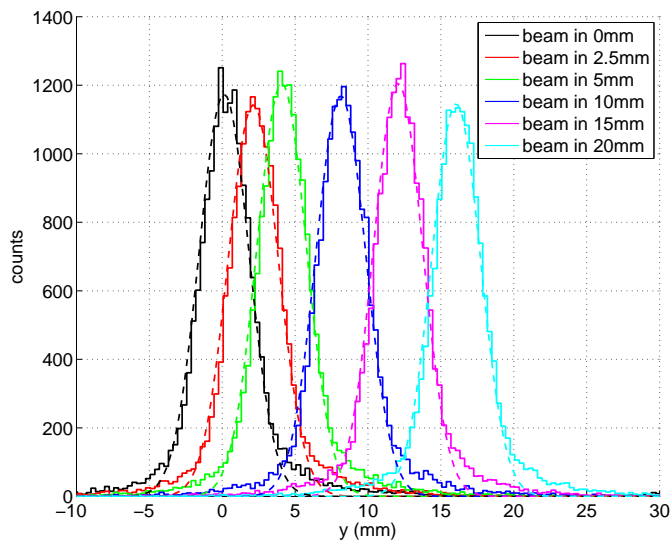}
\caption{\footnotesize Fine beam neutron scan along $y$. The spatial
resolution is given by the FWHM and corresponds to $4\,mm$.}
\label{spaty8jndh}
\end{figure}

The spatial resolution was calculated as already shown for the
Multi-Blade version V1 according to \cite{patrickinforis}. By using
a very collimated beam, of about $0.2\times10\,mm^2$, we scan one
cassette and a half of the detector. Each step is $0.9\,mm$ along
the $x$-direction. The cassette 1 is from $x=0\,mm$ to $x=10\,mm$,
the cassette 2 starts at $x=10\,mm$. Figure \ref{figexp0n74jdfiwhud}
shows the reconstructed image and its projection on the
$x$-direction obtained by adding together all the images taken in
the scan.
\\ For each step either a wire or two are firing. A the switching
point between the two cassette we observe the drops in the counts.
\paragraph{Spatial resolution: x}
We quantify the spatial resolution along the $x$-direction in the
same way as for the version V1.
\\ We scan the detector surface to obtain the events distribution to
calculate the mutual information \cite{patrickinforis} which is
shown in Figure \ref{mutinforfgjk830nsbbcd}. We use the threshold of
$0.47\,bits$ which corresponds to the standard FWHM resolution
definition and we obtain a value of $3.16\,mm$.
\\ This value is slightly better of the one found for the version V1
because in the version V2 the gap between wires and converter is
diminished. It is the first part of the ionization path, on average,
that gives more signal and it is closer to the fragment emission
point, this improves the spatial resolution.
\\ Since the detector is inclined at
$5^{\circ}$ the actual spatial resolution is given by the
projection: $3.16\,mm\cdot\sin(5^{\circ})=0.275\,mm$.
\paragraph{Spatial resolution: y}
The spatial resolution given by the strips is also enhanced thanks
to the narrower gas gap.
\\ Figure \ref{spaty8jndh} shows a scan performed along $y$. The
spatial resolution is given by the FWHM and is $4\,mm$.
\subsubsection{Dead time}
\begin{figure}[!ht]
\centering
\includegraphics[width=10cm,angle=0,keepaspectratio]{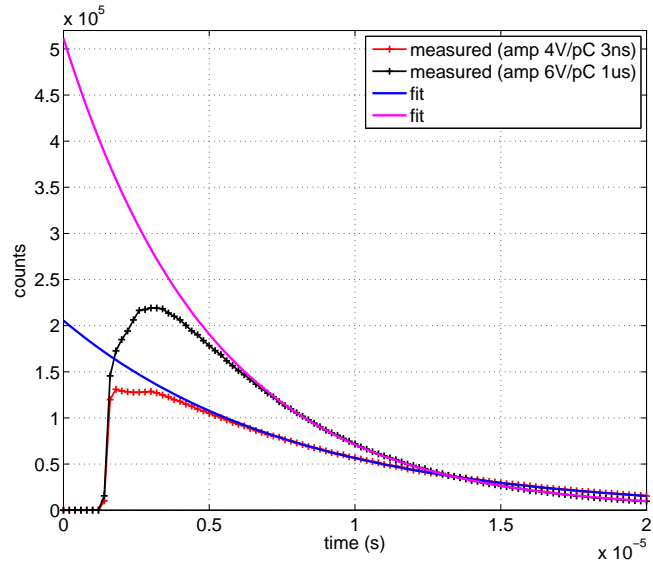}
\caption{\footnotesize Time distribution of neutron events recorded
with the Multi-Blade, the fit shows the theoretical behavior. Two
anode amplifiers have been used.} \label{dtmbv6scfa}
\end{figure}

The intrinsic dead time of a detector is due to its physical
characteristics; here we measure the entire dead time from the
detector to the end of the whole electronic chain.
\\ Neutrons arrive at the detector according to an exponential distribution
assuming the process to be Poissonian. A way to measure dead time is
to record the difference in the arrival time between two successive
neutrons on the detector; in principle their distribution should
follow an exponential:
\begin{equation}
f(t)=\frac1{\tau}e^{-t/\tau}
\end{equation}
where the time $\tau$ represents the average time is in between two
events; $\nu=1/\tau$ is the counting rate.
\\ In practice the detector is characterized by a dead time $t_D$
which is the minimum time interval that separates two correctly
recorded events. As a result the distribution measured with the
detector should move away from the theoretical behavior near and
below $t_D$. Moreover, if the detector is ideally non-paralyzable,
the distribution has to show a sharp cutoff at $t_D$ because the
probability of measuring an event between $t=0$ and $t=t_D$ is zero.
\\ In a paralyzable case the passage is smoother.
\\ Figure \ref{dtmbv6scfa} shows the measured times between
neutrons on the detector. The two anode outputs of a single cassette
where added and the resulting signal discriminated. The time between
every couple of discriminated events was recorded for $T=300\,s$.
\\ The measurement was performed by using two kind of amplifiers. One is
the standard Multi-Blade amplifier of $1\,\mu s$ shaping time and
the second is a fast amplifier with $3\,n s$ shaping time.
\\ The fits in Figure \ref{dtmbv6scfa} represent the theoretical
behavior. \\ The value for $\tau$, for both the measured
distributions, was obtained by the calculation of the maximum
likelihood estimator for the exponential distribution. It is:
\begin{equation}
\tau = \frac{\sum_i n_i (t_i-t_s)}{\sum_j n_j}
\end{equation}
where $t_s$ is the minimum time for which we consider the measured
distribution to behave as expected. We can assume that a time $t_s$
exists above which the measured distribution follows the exponential
behavior. We assume $t_s=7\,\mu s$.
\\ By knowing $\tau$ and the measurement duration $T=300\,s$, the
total number of neutrons that have generated a signal in the
detector but, due to dead time, have not all been recorded, is given
by $N_0 = \frac{T}{\tau}$.
If we integrate instead the measured distribution we obtain the
number of events recorded $N_m$. The dead time is simply given by:
\begin{equation}
t_D = \frac{N_0-N_m}{N_0}\tau = \frac{\frac{T}{\tau}-N_m}{T}\tau^2
\end{equation}
For the $1\mu s$ amplifier we obtain $t_D =
(1.58\pm0.08)\cdot10^{-6}\,s$, the fast amplifier gives $t_D =
(1.5\pm0.1)\cdot10^{-6}\,s$.

\section{Conclusions}
The Multi-Blade detector is a promising alternative to
$^{3}He$-based detectors, to accomplish the high spatial resolution
and the high count rate capability needed in the new applications of
neutron reflectometry. Two prototypes have been developed and
assembled at ILL to show we can overcome the reasonable limits of
$^3He$-based detectors in terms of spatial resolution.
\\ Exploiting solid $^{10}B_4C$-layers operated in a proportional gas chamber,
it can attain a suitable detection efficiency for neutron
reflectometry applications. A single $^{10}B_4C$ conversion layer is
operated at grazing angle with respect to the incoming neutron
direction. This configuration has at least three advantages: the
detection efficiency is increased, the neutron flux is split between
many wires and, the most important, the spatial resolution improved.
\\ In order to get a suitable detection efficiency, compared to
$^3He$, the converter layers must be operated at an angle of
$10^{\circ}$ or below.
\\ We studied two approaches to be used in the Multi-Blade
implementation: either with one or with two converters. The latter
has more technical issues that makes its realization more difficult.
The single layer detector is finally the choice to make to keep the
mechanics reasonably simple. The extra advantage of having only one
converter is that the coating can be of any thickness above $3\,\mu
m$ without affecting the efficiency, while for the two layer option
its thickness should be chosen carefully. Moreover, in the two layer
configuration the substrate choice is also crucial because it should
be kept as thin as possible to avoid neutron scattering and this
leads to mechanical issues. In a single-layer detector it can be
integrated in the holding structure. Moreover, since the efficiency
is saturated as the thickness of the single-layer exceeds $3\,\mu
m$, we can make a coating a few hundreds of $\mu m$ thick. A thick
layer also functions as an integrated collimator. Any neutron that
comes from the sides of the detector has less probability to be
detected with respect to those impinging the detector from the
front. Neutron background is then decreased.
\\ We conceived a detector to be modular in order to be
versatile: it is composed of modules called \emph{cassettes}. We
operated the two Multi-Blade prototypes at either
$\theta=10^{\circ}$ and $\theta=5^{\circ}$. In each of the solutions
proposed for the cassette concept the read-out system has to be
crossed by neutrons before reaching the converter. The mechanical
challenge in the read-out system construction is to minimize the
amount of material on the neutron path to avoid scattering that can
cause misaddressed events in the detector. The choice fell on
polyimide substrates. It can be eventually replaced with more
suitable materials.
\\ The detector is operated at atmospheric pressure. This makes it
suitable to be operated in vacuum. Moreover, cost effective
materials can be used inside the detector because outgassing is not
an issue.
\\ Since the detector is modular the main issue is its uniformity.
In the presented prototype we got about $50\%$ drop in efficiency in
the overlap region between cassettes.
\\ The presented Multi-Blade showed a very high spatial resolution,
it was measured to be about $0.3\,mm$ in one direction and about
$4\,mm$ in the other one.
\\ We measured the neutron detection efficiency for both prototypes
at $2.5$\AA \, neutron wavelength. The first prototype has an
efficiency of about $28\%$ employing sputtered $^{10}B_4C$-layers
inclined at $10^{\circ}$. This result is in a perfect agreement with
the expected efficiency \cite{gregor}, \cite{fratheo}. Since the
efficiency, in the single layer option does not depend on the
converter thickness above $3\,\mu m$, in the second prototype we
investigated a different deposition method: a $^{10}B$ glue-based
painting. This thick painted layer functions also as an integrated
collimator inside the detector. The resistivity of the $^{10}B$
painting is larger than the sputtered $^{10}B_4C$-layers but it
seems to be acceptable and does not cause issues to the charge
evacuation. We measured the efficiency of the second prototype
operated at $5^{\circ}$ and we only got about $8\%$. The coarse
granularity of the painting makes the inclination effect vanish. The
expected efficiency for a sputtered layer of which the roughness is
widely below the $\mu m$ scale is about $43\%$ (at $2.5$\AA). There
is no reason not to get the calculated efficiency with a suitable
painting.
\\ We measured the detector dead time, including the read-out electronics, to
be about $1.5\,\mu s$.
\\ The single layer option represents a good candidate to push
the performances of $^3He$ detectors. Further studies need to
address the uniformity problems. If a simple coating technique is
found, e.g. painting containing grains of a suitable size that
assures a uniform layer, the Multi-Blade could be a cost-effective
and high performance alternative. Its production is easy to be
implemented.

\acknowledgments The author (F.P.) would like to thank the entire
neutron detector group (SDN) at ILL for the practical and
intellectual support given to develop the work contained in this
article. In particular a special thank goes to B. Gu\'erard, the
detector group head at ILL, who first had the idea of the
Multi-Blade detector \cite{buff1} who gave the possibility to the
author to work on this subject. Specifically we would like to thank
J.C. Buffet and Q. La Manna for the mechanical designs, S. Cuccaro
for the mechanical support, J.F. Clergeau and J.M. Rigal for the
help in developing the electronics.
\\ \noindent The authors want also to thank the Thin Film Physics Division of
Link\"{o}ping University (Sweden) - especially C. H\"{o}glund, - for
the Boron Carbide coatings.


\begin{thebibliography}{9}

\bibitem{esstdr} S. Peggs, \emph{ESS Technical Design Report}, ESS-doc-274, 23 April 2013.
ISBN 978-91-980173-2-8,
\href{http://eval.esss.lu.se/cgi-bin/public/DocDB/ShowDocument?docid=274}{http://eval.esss.lu.se/cgi-bin/public/DocDB/ShowDocument?docid=274}.

\bibitem{gebauer1} B. Gebauer et al., \emph{Towards detectors for next generation spallation neutron sources},
Proceedings of the 10th International Vienna Conference on
Instrumentation, Nuclear Instruments and Methods in Physics Research
Section A: Accelerators, Spectrometers, Detectors and Associated
Equipment, Volume 535, Issues 1-2, 2004, Pages 65-78,
\href{http://dx.doi.org/10.1016/j.nima.2004.07.266}{0168-9002,
10.1016/j.nima.2004.07.266}.

\bibitem{figaro}
R. A. Campbell et al., \emph{FIGARO: The new horizontal neutron
reflectometer at the ILL}, The European Physical Journal Plus,
Springer-Verlag, Volume 126, Issue 11, 2011, Pages 1-22,
\href{http://dx.doi.org/10.1140/epjp/i2011-11107-8}{10.1140/epjp/i2011-11107-8}.

\bibitem{cubittD17}
R. Cubitt et al., \emph{D17: the new reflectometer at the ILL},
Journal of Applied Physics A, Springer-Verlag, Volume 74, Issues 1,
1 December 2002, Pages s329-s331,
\href{http://dx.doi.org/10.1007/s003390201611}{10.1007/s003390201611}.

\bibitem{rainbow1} J. Stahn et al., \emph{Focusing specular neutron reflectometry for small samples}.
Eur. Phys. J. Appl. Phys., Volume 58, 2012,
\href{http://www.epjap.org/article_S1286004212202952}{10.1051/epjap/2012110295}.

\bibitem{cubitt2}
R. Cubitt et al., \emph{Neutron reflectometry by refractive
encoding}, The European Physical Journal Plus, Springer-Verlag,
Volume 126, Issues 11, 11 November 2011, Pages 1-5,
\href{http://dx.doi.org/10.1140/epjp/i2011-11111-0}{10.1140/epjp/i2011-11111-0}.

\bibitem{rainbow2} R. Cubitt et al., \emph{Refraction as a means of encoding wavelength for neutron reflectometry}.
Nuclear Instruments and Methods in Physics Research Section A:
Accelerators, Spectrometers, Detectors and Associated Equipment,
Volume 558, 2006, Pages 547-550,
\href{http://dx.doi.org/10.1016/j.nima.2005.12.045}{10.1016/j.nima.2005.12.045}.

\bibitem{buff1} J.C. Buffet et al., \emph{Advances in detectors for single crystal
neutron diffraction}, Nuclear Instruments and Methods in Physics
Research Section A: Accelerators, Spectrometers, Detectors and
Associated Equipment, Volume 554, Issues 1-3, 1 December 2005, Pages
392-405, ISSN 0168-9002,
\href{http://dx.doi.org/10.1016/j.nima.2005.08.018}{10.1016/j.nima.2005.08.018}.

\bibitem{buff3} J.C. Buffet et al., \emph{Study of a 10B-based Multi-Blade
detector for Neutron Scattering Science}, Nuclear Science Symposium
and Medical Imaging Conference (NSS/MIC), Transaction Nuclear
Science Conference Record IEEE - Anaheim, 2012, PAGES 171-175, ISSN
1082-3654,
\href{http://ieeexplore.ieee.org/stamp/stamp.jsp?tp=&arnumber=6551086&isnumber=6551044}{10.1109/NSSMIC.2012.6551086}.

\bibitem{jonisorma}
J. Birch et al., \emph{$^{10}B_4C$ Multi-Grid as an Alternative to
$^{3}He$ for large area neutron detectors}, IEEE T. Nucl. Sci.,
Volume PP, Issue 99, 17 January 2013, Pages 1-8, ISSN 0018-9499,
\href{http://dx.doi.org/10.1109/TNS.2012.2227798}{10.1109/TNS.2012.2227798}.

\bibitem{kleinjalousie} M. Henske et al., \emph{The 10B based Jalousie neutron detector $-$ An alternative
for 3He filled position sensitive counter tubes}, Nucl. Instrum.
Meth. A, Volume 686, 11 September 2012, Pages 151-155, ISSN
0168-9002, \href{http://dx.doi.org/10.1016/j.nima.2012.05.075}
{10.1016/j.nima.2012.05.075}.

\bibitem{gregor}
D. S. McGregor et al., \emph{Design considerations for thin film
coated semiconductor thermal neutron detectors$-I$: basics regarding
alpha particle emitting neutron reactive films}, Nuclear Instruments
and Methods in Physics Research Section A: Accelerators,
Spectrometers, Detectors and Associated Equipment, Volume 500,
Issues 1-3, 11 March 2003, Pages 272-308, ISSN 0168-9002,
\href{http://dx.doi.org/10.1016/S0168-9002(02)02078-8}{10.1016/S0168-9002(02)02078-8}.

\bibitem{fratheo} F. Piscitelli et al., \emph{Analytical modeling
of thin film neutron converters and its application to thermal
neutron gas detectors}, Journal of Instrumentation, Volume 8,
P04020, April 2013,
\href{http://iopscience.iop.org/1748-0221/8/04/P04020/}{10.1088/1748-0221/8/04/P04020}.

\bibitem{carina}
C. H\"{o}glund et al., \emph{$B_4C$ thin films for neutron
detection}, J. Appl. Phys., Volume 111, Issue 10, 23 May 2012, Pages
10490-8, ISSN 0168-9002,
\href{http://link.aip.org/link/?JAP/111/104908/1}{10.1063/1.4718573}.

\bibitem{patrickinforis} P. Van Esch et al., \emph{An information-theoretical approach
to image resolution applied to neutron imaging detectors based upon
discriminator signals}, Proceeding of ANNIMA, 2013,
\href{http://arxiv.org/abs/1307.7507}{arXiv:1307.7507}.

\end{thebibliography}
\end{document}